# Trace element geochemistry of ordinary chondrite chondrules: the type I/type II chondrule dichotomy

Emmanuel Jacquet[1,2], Olivier Alard[3], Matthieu Gounelle[2,4]

[1]Canadian Institute for Theoretical Astrophysics, 60 St George Street, Toronto, ON, M5S 3H8, Canada.

[2]Institut de Minéralogie, de Physique des Matériaux et de Cosmochimie, CNRS & Muséum National d'Histoire Naturelle, UMR 7590, 57 rue Cuvier, 75005 Paris, France.

[3]Géosciences Montpellier, UMR 5243, Université de Montpellier II, Place E. Bataillon, 34095 Montpellier cedex 5, France.

[4]Institut Universitaire de France, Maison des Universités, 103 boulevard Saint-Michel, 75005 Paris, France.

E-mail: emjacquet@mnhn.fr

## *Abstract*

We report trace element concentrations of silicate phases in chondrules from LL3 ordinary chondrites Bishunpur and Semarkona. Results are similar to previously reported data for carbonaceous chondrites, with rare earth element (REE) concentrations increasing in the sequence olivine < pyroxene < mesostasis, and heavy REE (HREE) being enriched by 1-2 orders of magnitude (CI-normalized) relative to light REE (LREE) in ferromagnesian silicates, although no single olivine with very large LREE/HREE fractionation has been found. On average, olivine in type II chondrules is poorer in refractory lithophile incompatible elements (such as REE) than its type I counterpart by a factor of ~2. This suggests that olivine in type I and II chondrules formed by batch and fractional crystallization, respectively, implying that type II chondrules formed under faster cooling rates (> ~ 10 K/h) than type I chondrules. Appreciable Na concentrations (3-221 ppm) are measured in olivine from both chondrule types; type II chondrules seem to have behaved as closed systems, which may require chondrule formation in the vicinity of protoplanets or planetesimals. At any rate, higher solid concentrations in type II chondrule forming regions may explain the higher oxygen fugacities they record compared to type I chondrules. Type I and type II chondrules formed in different environments and the correlation between high solid concentrations and/or oxygen fugacities with rapid cooling rates is a key constraint that chondrule formation models must account for.

## *1. Introduction*

While the vast majority of primitive meteorites contain chondrules — millimeter-sized silicate spherules presumably formed by solidification of once-molten droplets —, how these formed is still essentially unknown, however critical such knowledge would be for the understanding of the early solar system. In recent decades, attention has mostly focused on "nebular scenarios" based e.g. on shock waves (Desch et al. 2005), lightning (Desch & Cuzzi 2000), winds (Shu et al. 2001; Salmeron & Ireland 2012) or magnetohydrodynamic turbulence (McNally et al. 2013). This may be traced to arguments against "planetary processes" summarized by Taylor et al. (1983) e.g. old chondrule ages, unfractionated compositions, etc. However, evidence is also mounting that chondrule-forming environments were quite different from standard "average" protoplanetary disks. In particular, solid/gas ratios orders of magnitude above solar seem required by the frequency of compound chondrules (Gooding & Keil 1981), the high FeO contents of many chondrules (Grossman et al. 2012) and the retention of Na in chondrules (Alexander et al. 2008; Hewins et al. 2012). This has reawakened interest in planetary scenarios such as the planetesimal splash model (Sanders & Scott 2012; Asphaug et al. 2011) or impact-generated plume environments (Symes et al. 1998; Fedkin & Grossman 2013). Chondrule formation is obviously not understood yet, and the body of cosmochemical constraints itself is still in flux.

Chondrules show considerable chemical, isotopic and petrographic diversity (Grossman & Wasson 1983; Jones 2012) indicating that chondrule formation took place in a number of distinct reservoirs, under different conditions, a variety that must be accounted for by any successful chondrule-forming theory. In particular, ferromagnesian chondrules (which make up the majority of chondrules) can be classified to first order in type I and type II chondrules (e.g. Scott & Taylor 1983). The former have little FeO, significant amounts of metallic iron-nickel, and are somewhat depleted in volatile elements relative to solar while the latter have more ferroan silicates, overall higher total iron content (Huang et al. 1996), and more chondritic abundances of volatile elements. What this dichotomy means is still unclear: type II chondrules may have formed in regions with higher overall oxygen fugacities, e.g. through higher dust/gas ratios (Schrader et al. 2011), or owe their redox state to their own composition e.g. carbon content (Connolly et al. 1994), or yet differ from possibly more primitive type I chondrules only in the amount of iron to be oxidized and/or the duration of the thermal episode time (Villeneuve 2010).

To make progress on these issues, trace element geochemistry is a powerful tool. Indeed, the partitioning of the different trace elements among the different phases, in particular silicates, of chondrules, is very sensitive to the formation conditions, and this at the order-of-magnitude level. While trace element bulk compositions of chondrules has been determined for a few decades already (e.g. Grossman and Wasson 1983; Misawa & Nakamura 1988; Pack et al. 2004; Pack et al. 2007; Inoue et al. 2009; Gordon 2009), Secondary Ion Mass Spectrometry (SIMS) has been used from the 1990s onward to investigate their concentration at the mineral scale, especially in ordinary chondrites (e.g. Alexander 1994; Jones & Layne 1997; Ruzicka et al. 2008). However, the very low abundances of many trace elements of interest in ferromagnesian silicates, in particular rare earth elements (REE), are often below detection in SIMS. With its lower detection limits, Laser Ablation Inductively Coupled Plasma Mass Spectrometry (LA-ICP-MS) is an advantageous alternative technique, which also allows analyses to be carried out at a greater pace. We (Jacquet et al. 2012) recently conducted *in situ* LA-ICP-MS analyses of chondrules, mostly of type I, in the carbonaceous chondrites Vigarano (CV3), Renazzo (CR2) and Acfer 187 (CR2). The relatively high incompatible element content of olivine was found consistent with batch crystallization, and thence cooling rates slower than ~10 K/h. However, light REE abundances exceeding equilibrium partitioning predictions also indicated kinetic effects, especially for low-Ca pyroxene whose crystallographic structure appears to record cooling rates of order 1000 K/h. Unless pyroxene crystallization was an event distinct from olivine formation, this appeared to suggest an *accelerating* cooling rate after chondrule heating, quite distinct from the usual picture of chondrule thermal history (Hewins et al. 2005). Yet Villeneuve (2010) had advocated a similar temperature curve for type II chondrules and the composition of CR chondrite metal was also found to be consistent with this picture (Jacquet et al. 2013). It is important to keep in mind that estimated cooling rates depend on the assumed form of the cooling law: yet another distinct possibility proposed by Wasson (1996); Rubin and Wasson (2003) to prevent loss of sodium, is that chondrules result from a *series* of several short heating events.

In this paper, we extend our study to non-carbonaceous chondrites, and specifically LL3 ordinary chondrites Bishunpur and Semarkona. Ordinary chondrites contain many more type II chondrules than carbonaceous chondrites (in which only 3 type II chondrules were analyzed in our previous study (Jacquet et al. 2012)) and will allow insight on the type I/type II chondrule dichotomy. Although, as mentioned above, SIMS-based analyses of ordinary chondrite chondrule silicates have already been reported (Alexander et al. (1994); Jones and Layne (1997); Ruzicka et al. (2008)), the lower detection limits of LA-ICP-MS will allow deeper insight in the trace element budget of these chondrules, especially for REE, and this in a consistent fashion with our work on carbonaceous chondrites.

## *2. Samples and analytical procedures*

Two polished sections from the meteorite collection of the Muséum National d'Histoire Naturelle (MNHN; Paris) were used in this study: section Sp4 of the Bishunpur (LL3.1) and section ns1 of Semarkona (LL3.0), both observed falls. Analytical procedures were similar to those Jacquet et al.

(2012) and are only briefly summarized here. The sections were examined in optical and scanning electron microscopy (SEM—here a JEOL JSM-840A instrument). X-ray maps (which could not be performed on the Semarkona section) allowed apparent mineral modes of chondrules to be calculated using the JMicrovision software (www.jmicrovision.com). Minor and major element concentrations of documented chondrules were obtained with a Cameca SX-100 electron microprobe (EMP) at the Centre de Microanalyse de Paris VI (CAMPARIS).

Trace element analyses of selected chondrules were performed by LA-ICP-MS at the University of Montpellier II. The laser ablation system was a GeoLas Q$^+$ platform with an Excimer CompEx 102 laser and was coupled to a ThermoFinnigan Element XR mass spectrometer. The ICP-MS was operated at 1350 W and tuned daily to produce maximum sensitivity for the medium and high masses, while keeping the oxide production rate low ($^{248}$ThO/$^{232}$Th ≤ 1%). Ablations were performed in pure He-atmosphere (0.65 ± 0.05 L•min$^{-1}$) mixed before entering the torch with a flow of Ar (≈ 1.00 ± 0.05 L•min$^{-1}$). Laser ablation condition were: fluences ca. 12 J/cm² with pulse frequencies between 5 and 10 Hz were used and spot sizes of 26-102 µm, 51 µm being a typical value. With such energy fluences and pulse frequencies, depth speed for silicates is about 1 µm•s$^{-1}$. Each analysis consisted of 4 min of background analyses (laser off) and 40 s of ablation (laser on). Data reduction was carried out using the GLITTER software (Griffin et al. 2008). Internal standard was Si (except for augite, where Ca was preferred), known from EMP analyses. The NIST 612 glass (Pearce et al. 1997) was used as an external standard. The absence of contamination by other phases was checked by examination of the time-resolved GLITTER signal, SEM imaging of ablation craters and comparison with EMP data. Examples of ablation craters are shown in the Electronic Annex. For each chondrule, a geometric (rather than arithmetic) averaging was used to calculate mean concentrations in olivine and pyroxene, this in order to further minimize the impact of possible contamination by incompatible element-rich phases. In the following text and in the figures, and unless otherwise noted, the data reported will be *chondrule means*, that is, for each phase, the average of the different analyses performed on that phase in a given chondrule.

For chondrules where olivine, low-Ca pyroxene and mesostasis trace element analyses were all available (or if one of the first two, although not analysed, had a modal abundance less than 5 %), a reconstructed bulk chondrule composition (for non siderophile elements) has been calculated based on the modal data. Given in particular the uncertainties e.g. in the true (tri-dimensional) modes of the phases (see e.g. Hezel & Kießwetter 2010), those calculations should not be granted more than qualitative value.

# *3. Results*

## 3.1 Petrography and mineralogy

In this section we briefly describe the petrography of analyzed chondrules, some images of which are provided in Fig. 1. As in Jacquet et al. (2012), we call "chondrules" all silicate objects with petrographic evidence of melting regardless of their overall shape. Chondrule names begin with "Bi" or "Smk" depending on whether they are drawn from Bishunpur or Semarkona, respectively. Here, "type I" refers to low-FeO (less than 10 mol% fayalite in olivine) chondrules and "type II" to high-FeO ones (regardless of texture, unlike earlier usage, e.g. McSween (1977)), to which a letter A or B may be added to indicate olivine or pyroxene dominance (Ol/(Ol+Px)>90 vol% and < 10 vol%, respectively), or AB in the intermediate case. As to petrographic taxonomy *stricto sensu*, porphyritic, radial and barred types are denoted "P", "R" and "B" respectively, (to which "O", "OP", or "P" may be added to indicate relative proportions of olivine and pyroxene according to the same criteria as above). Finally, note that we will restrict usage of the name "enstatite" to low-Ca pyroxene with less than 10 mol% ferrosilite.

In Bishunpur, 10 type I and 9 type II chondrules, most porphyritic textured (but including one BO and one RP), were analyzed. Type I chondrules (e.g. Fig. 1a-e) display the typical mineralogical zoning of olivine crystals dominating the interior and enstatite laths concentrated around the margin. Type I chondrules are commonly surrounded by a fine-grained FeO-rich rim, and tend to show somewhat irregular outlines which may be interpreted as adhesions of microchondrules. Indeed, some microchondrules are occasionally found entirely trapped in the rim, although only one very microchondrule-rich object, Bi41 (Fig. 1d, e), reminiscent of those described by Rubin et al. (1982) and Krot et al. (1997) was found in this study [electron microprobe analyses of those are listed in Table EA6]. Most type IA or IAB chondrules contain some dusty olivine (i.e. originally ferroan relict olivine grains having undergone reduction and speckled with iron blebs) and contain Fe-Ni metal and troilite (the latter often overgrowing the former). Mesostasis is typically finely devitrified, sometimes with evidence of partial hydration (e.g. Rubin 2013), and augite overgrowths on enstatite laths are common. Enstatite laths, often poikilitically enclosing olivine, contain shrinkage cracks (Brearley & Jones 1998) generally devoid of the whitish fillings visible in CR and CV chondrite enstatite (see e.g. Fig. 3b of Marrocchi and Libourel (2013)). No granoblastic olivine aggregate such as those described by Libourel & Krot (2007) in CV chondrites was encountered.

Type II chondrules (e.g. Fig. 1h-k) contain normally zoned ferroan olivine and pyroxene with no apparent mineralogical zoning, except insofar as sulphides — their dominant opaques — tend to concentrate around the margins, sometimes in the form of a continuous silicate/sulphide intergrowth. Mesostasis appears more glassy than its type I counterpart, with varying abundances of augite crystallites. Qualitative X-ray maps for Na uncommonly reveal chondrule-wide zoning of mesostasis (for all chondrule types studied), consistent with the 15 % frequency of such features seen for type I chondrules by Grossman et al. (2002); those might be due to aqueous alteration. Forsteritic relicts are fairly uncommon. A few type II chondrules consist of massive low-Ca pyroxene poikilitically enclosing rounded olivine chadacrysts, with some anorthitic plagioclase and no glassy mesostasis, one of which was analyzed in this study (Bi7; Fig. 1j). A 4 mm-diameter ferroan olivine-dominated object was included in the analysed suite (Bi1; Fig. 1h).

In Semarkona, 5 type I and 3 type II chondrules were analysed (Fig. 1f,g,l). Individual chondrule petrography is similar to the above but it should be commented that the texture of Semarkona differs from that Bishunpur (at least for the sections studied) in having more tightly packed and mutually indented chondrules (see Fig. 1m,n). That is, Semarkona is a "cluster chondrite" as described by Metzler (2012) rather than a "normal" clastic-textured one. Suphide veneers are also more frequent in Semarkona and less biased toward type II chondrules than in Bishunpur.

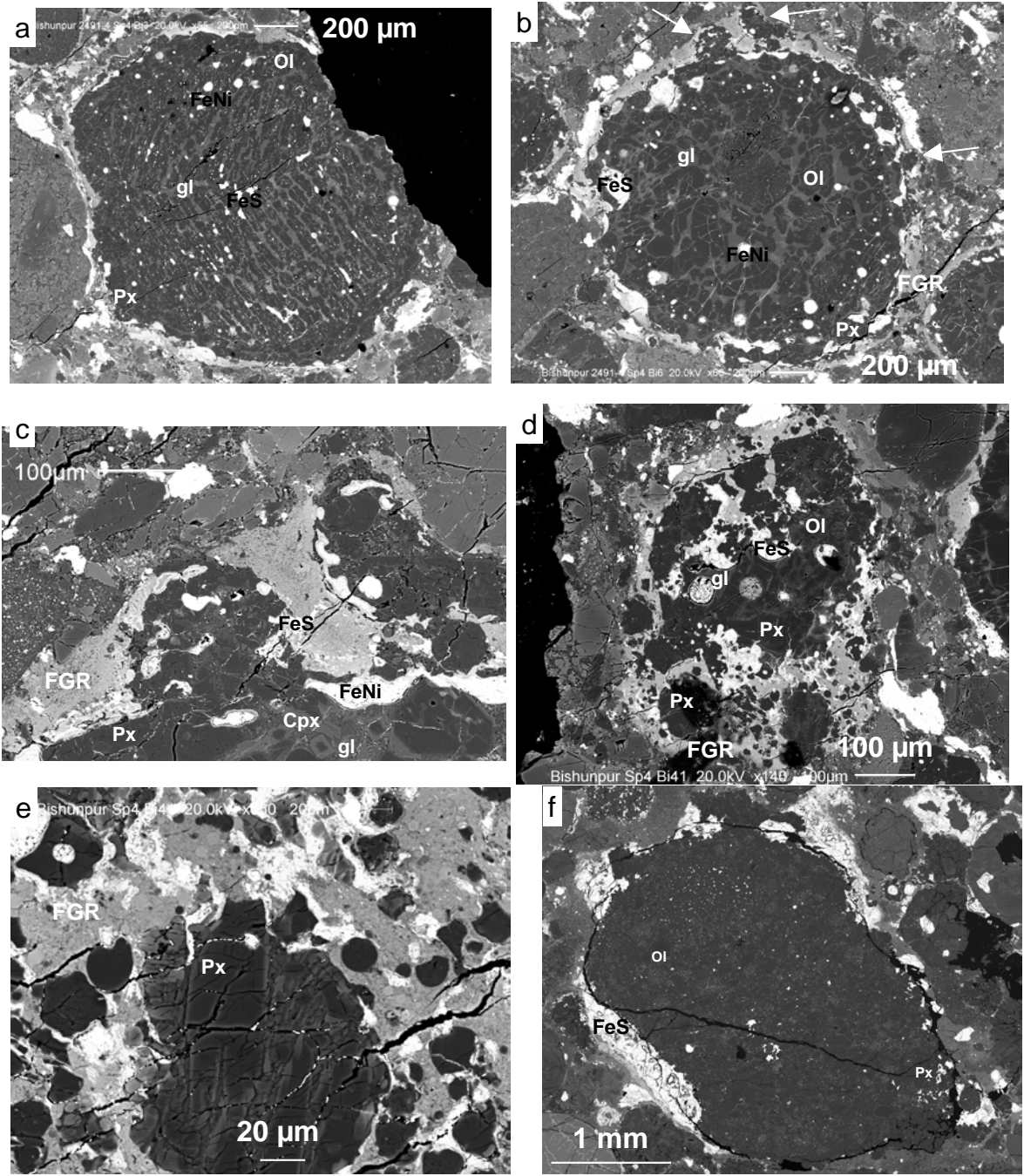

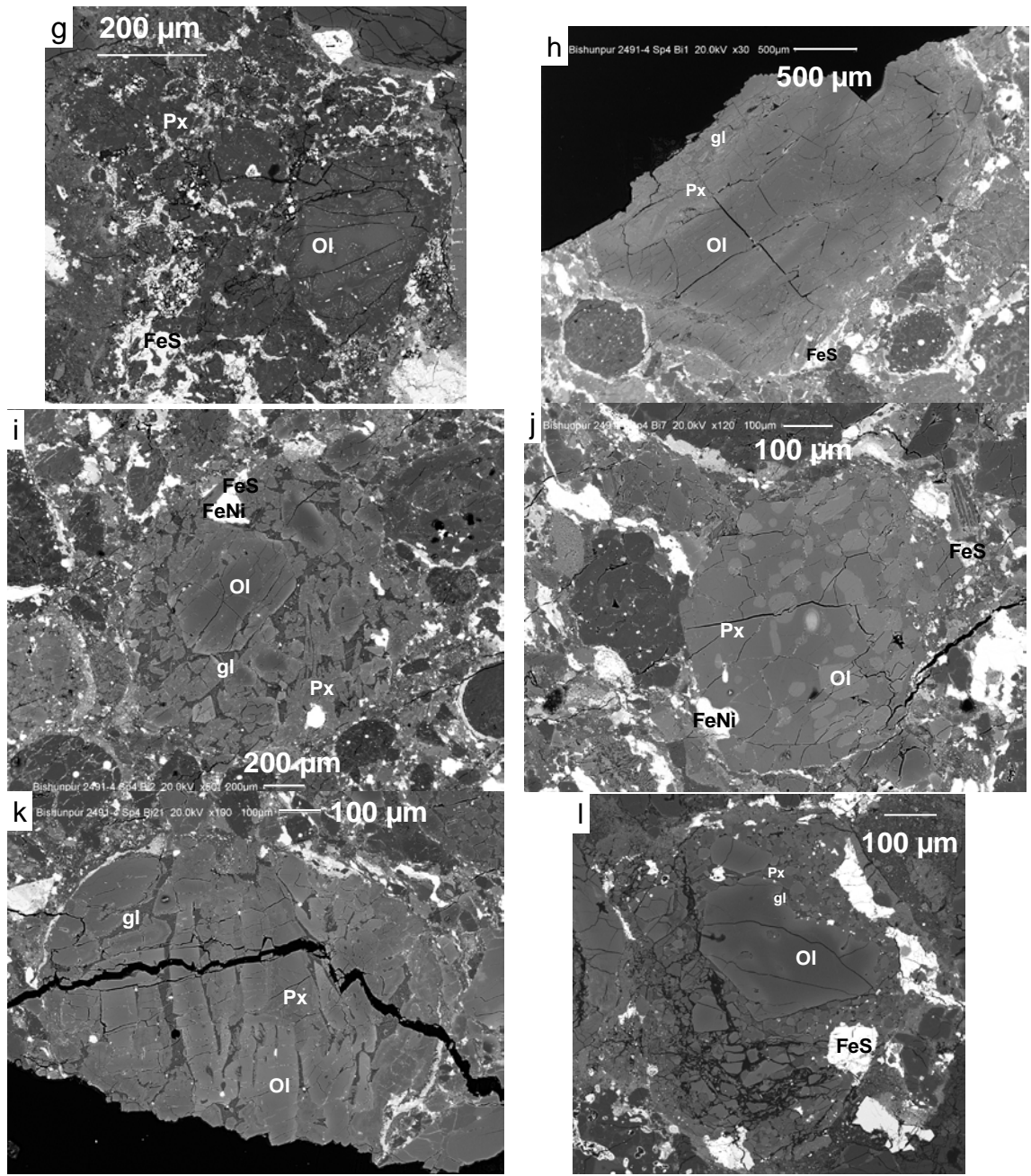

**Figure 1**: Back-scattered electron images of selected chondrules and other inclusions analyzed in this study ("Bi" prefix refers to Bishunpur chondrules and "Smk" to Semarkona chondrules). (a) Barred olivine type I chondrule Bi3. Olivine bars are mostly dusty and exhibit two orientations in the upper and lower parts of the chondrule, suggesting, consistent with the overall shape, that it may actually be compound. Fe-Ni nodules and patchy sulfide occur in olivine and the finely devitrified mesostasis. An enstatite layer is in contact with a ferroan fine-grained rim. (b) Type IAB porphyritic chondrule Bi6, with both "clear" and "dusty" olivine crystals embedded in devitrified mesostasis. Nodules of Fe-Ni, included in amoeboid troilite, are concentrated in the enstatite-rich margin. The surface of the chondrule is irregular with a few apparent adhesions of smaller (< 100 µm) chondrules (some designated by the white arrows) all embedded in a ferroan fine-grained rim. (c) Close-up of microchondrules-bearing upper rim of same. (d) Porphyritic, pyroxene-dominated type IAB chondrule Bi41, of irregular shape, with enstatite crystals bordered by augite and an olivine grain

cluster on the right. A thick ferroan fine-grained rim with numerous enstatite microchondrules (the largest with radial/barred texture), sometimes adhering to the largest object, surrounds the chondrule and fills in many embayments in it. Fayalite microchondrules have also been found. Troilite (enclosing remnant Fe-Ni metal) occurs as nodules in the chondrule and also impregnates the rim. The whole object could be a clast from a preexisting parent body. (e) Close-up pictures of the fine-grained rim and microchondrules, sometimes compound or cratered, around same. (f) Type I porphyritic macrochondrule Smk1, 2x3 mm, with equant crystals of olivine (both dusty and clear) and enstatite set in a devritrified mesostasis. A thick albeit discontinuous sulfide veneer borders the object. (g) Chondrule Smk24, with coarse ferroan olivine partially reduced on the outside. Enstatite (with olivine chadacrysts) and sulfide dominate the irregularly-shaped object, with metal specks and mesostasis patches also present. (h) Object Bi1, 4 mm across, essentially of monosomatic olivine ($Fo_{86-91}$), with elongate devitrified mesostasis patches bordered by zoned low-Ca pyroxene (bronzite) with augite overgrowth. (i) Type IIAB porphyritic chondrule Bi2, of oval shape with large (700 µm) olivine phenocryst. Low-Ca pyroxene (hypersthene) occurs as laths or oikocrysts around olivine grains, with interstices filled by glassy mesostasis, with occasional crystallite clusters. Fe-Ni metal is more or less corroded by troilite. (j) Type IIAB chondrule fragment Bi7 with hypersthene (slightly enriched in Ca and Al toward the right-hand side) poikilitically enclosing rounded olivine grains. Fe-Ni metal is visible in the olivine-free area of the pyroxene. The texture is reminiscent of chondrule R42 in Renazzo (Fig. 1i of Jacquet et al. 2012), although the oikocryst was augitic there. (k) Type IIAB BOP/POP chondrule Bi21, ovoid in shape, with subvertical bars of hopper-textured olivine transecting often zoned (sometimes in an oscillatory fashion) bronzite crystals, with augite overgrowths sometimes protruding in the mesostasis. (l) Type II porphyritic chondrule Smk15, dominated by a 300 µm phenocryst (from which the olivine analyses were taken), included in a groundmass of olivine, low-Ca pyroxene and mesostasis, with some peripheral sulfide. Abbreviations: Ol = olivine, Px = low-Ca pyroxene, Cpx = Ca-rich pyroxene, gl = mesostasis, FeNi = Fe-Ni metal, FeS = troilite, FGR = fine-grained rim.

m

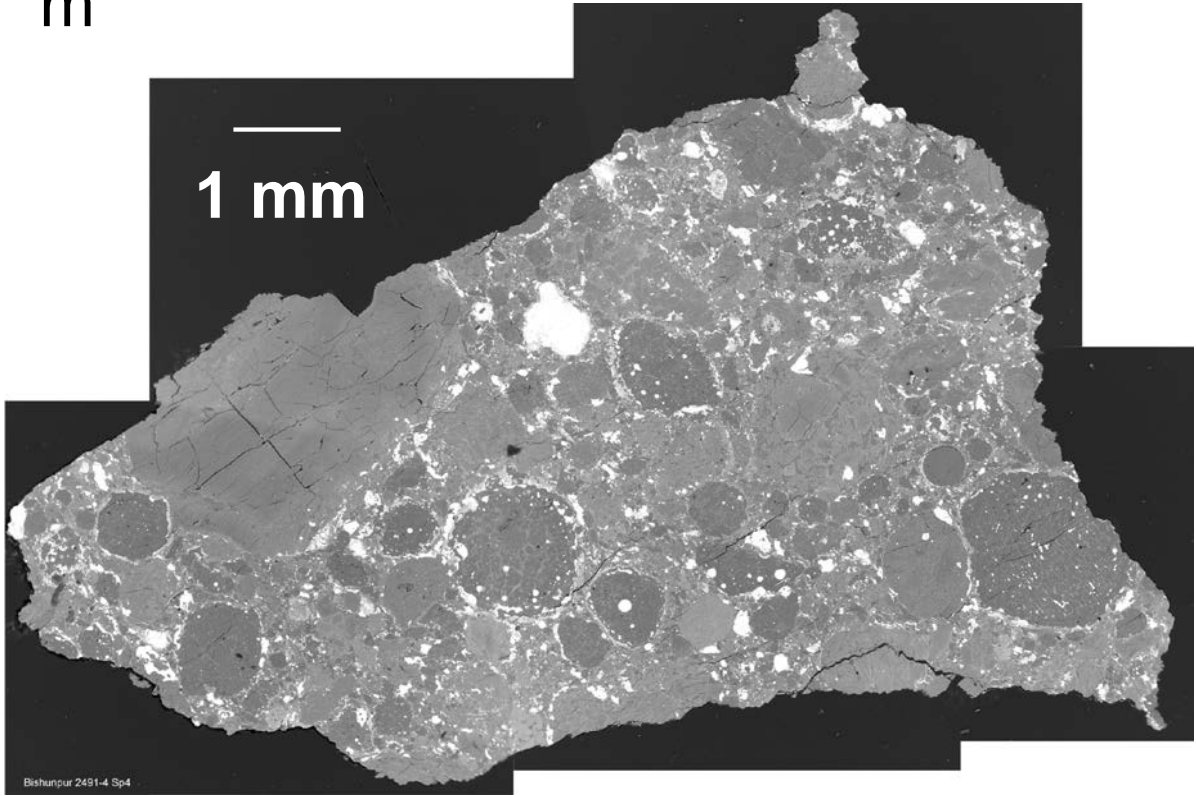

n

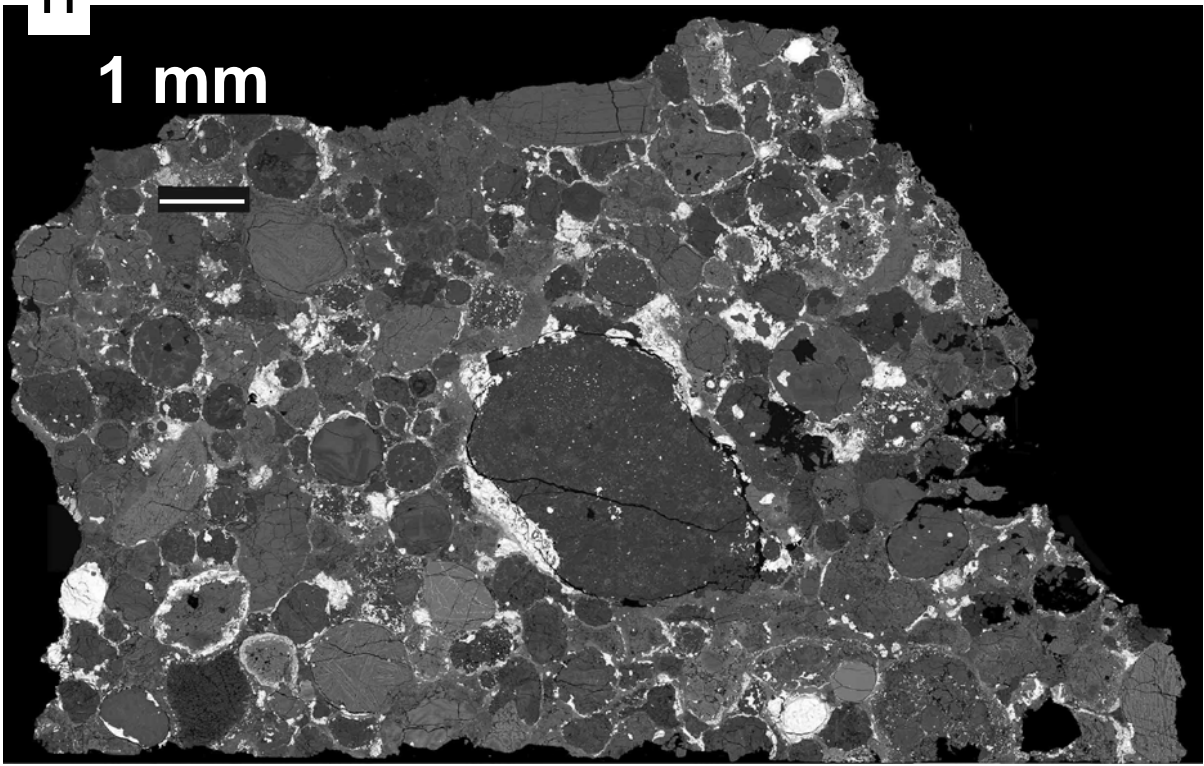

**Figure 1 (Continued)**: (m) Back-scattered electron mosaic of Bishunpur section Sp4. (n) Back-scattered electron mosaic of Semarkona section ns1. Deformation of close-packed chondrules as well as abundant troilite veneers are apparent.

### 3.2 Chemistry

Trace element data for each chondrite and phase are summarized in Tables 1-3, and individual chondrule data are provided in the Electronic Annex. Rare Earth Element and other lithophile element patterns are shown in Fig. 2 (with comparison to Vigarano; results for Renazzo are not shown for clarity but are very similar to the latter, cf Jacquet et al. 2012). We now describe the chemical composition of the chondrules, with particular emphasis on Rare Earth Elements (REE). We denote by LREE and HREE the light and heavy REE, respectively. Normalizations to CI chondritic abundances will be made using the Lodders (2003) values and denoted by the subscript "N". Since our few analyses of Semarkona chondrules appear to agree with those of Bishunpur, we shall discuss results from both meteorites collectively but distinguish between type I and type II chondrules.

Mesostases exhibit flat REE patterns with generally a weak LREE enrichment over HREE ($0.5 \leq (Ce/Yb)_N \leq 2.0$). La concentrations span 5-21 x CI. Weak Eu anomalies occur (Eu/Eu* = 0.1-1.3 where $Eu^*_N = (Sm_N + Gd_N)/2$). This agrees well with SIMS measurements by Alexander et al. (1994), Jones and Layne (1997) and Ruzicka et al. (2008). Mesostases in type I chondrules have refractory lithophile element contents comparable, if slightly higher on average, to their type II counterparts but are distinctly depleted in moderately volatile elements (Fig. 2), e.g. P or Na (although one type I chondrule, Bi41 (Fig. 1g), has P concentration of 8 x CI in its mesostasis).

Two augite crystals were analyzed in type I chondrules in Bishunpur. They show HREE enrichments relative to LREE (with a flat HREE segment slightly above mesostasis levels) and pronounced negative Eu anomalies (Eu/Eu* = 0.1 and 0.2), in agreement with Alexander et al. (1994) and Jones and Layne (1997).

Low-Ca pyroxene is poorer in REE than augite, with La at 0.005-0.2 x CI and Lu at 0.08-1 x CI. The CI-normalized pattern steadily decreases from HREE to LREE. Compared to Vigarano and Renazzo (Jacquet et al. 2012), there is little flattening for the LREE ($0.5 \leq (La/Sm)_N \leq 1.3$ excepting Bi7 (0.1); $0.08 \leq (Ce/Yb)_N \leq 0.8$) and the positive Eu anomalies found in Vigarano are almost absent. Perhaps the latter were related to the whitish fillings of shrinkage cracks in Vigarano enstatite and/or to aqueous alteration or fine-scale co-crystallization of sulfides (Marrocchi and Libourel 2013, see also Kong et al. (2000)). The data are consistent with the combined range of those of Alexander (1994) and Jones and Layne (1997).

Olivine has generally the lowest REE abundances, with La at 0.008-0.2 x CI (excepting Bi9 ($Fo_{95}$) at 0.0009 x CI) and 0.004-0.02 x CI for type I and Lu at 0.04-0.7 x CI and 0.03-0.2 x CI for type II chondrules, respectively. Type I chondrule olivine is on average richer in refractory lithophile elements (by a factor of ~2) than type II (Fig. 3c; 4a,c). Alexander (1994) had also found that olivine was enriched in incompatible elements for forsterite contents above 98 mol%, which encompasses most type I chondrules in unequilibrated ordinary chondrites (and indeed, olivine from relatively FeO-rich type I chondrule Bi9 olivine, at $Fo_{93}$, has low $Lu_N$ = 0.04). This is also what we had observed for the 3 type II chondrules we analyzed in carbonaceous chondrites (Jacquet et al. 2012). We do not however confirm the high LREE contents ($La_N$, $Ce_N$ > ~0.1) reported by Alexander et al. (1994) and Ruzicka et al. (2008) (see e.g. Fig. 4c for Ce). As it appears as unlikely to us as to them that contamination e.g. by glass pockets systematically compromised their analyses (which had smaller spot sizes than ours), with their reported Al and Ca values comparable to ours (e.g. Fig. 3c), we suspect that this may be traced back to the higher detection limits of ion microprobes compared to LA-ICP-MS (and indeed, reported La concentrations reported by Alexander et al. (1994) are rarely above 3 times the stated standard deviation). Whatever that may be, the systematic depletion of type II chondrule olivine in LREE relative to type I's that we measure (Fig. 2a; Fig. 4c) gives us further confidence that we are adequately assessing the true average level of LREE abundances rather than background noise. Globally, the REE pattern roughly parallels that of low-Ca pyroxene, with flattening in the LREE segment (0.03 ≤ (La/Sm)$_N$ ≤ 1.6). Unlike the carbonaceous chondrites we have studied previously (Jacquet et al. 2012), no strong LREE/HREE fractionation is ever encountered (0.02 ≤ (Ce/Yb)$_N$ ≤ 0.7). Na concentrations (3-207 ppm) are consistent with EMP data of Hewins et al. (2012) though with very Na-depleted values in two type II chondrules (Smk15 and Smk20 with <1.2 and 3.1 ppm respectively) in Semarkona, in both case in relatively coarse phenocrysts (Fig. 1l). Na does not correlate with Mn or Ti which are mutually anticorrelated, but shows a diffuse positive trend with Fe (Fig. 3a, 4d).

The reconstructed bulk compositions show generally unfractionated refractory lithophile elements, with e.g. La varying between 0.6 and 3 x CI, with type I and type II chondrule averages being within error of each other (~1.5 x CI). Type I chondrules are however more depleted in moderately volatile elements than their type II counterparts. This is in agreement with bulk LA-ICP-MS analyses of coarse-grained porphyritic chondrules in Semarkona by Gordon et al. (2008) (see also Wasson et al. 2000). The most Si-rich type I chondrules are poorest in REE (Fig. 5a). Unlike type I chondrules, type II chondrules show a positive correlation between Na and Al at Na/Al = 0.8 (atomic; Fig. 5b), as seen previously e.g. by Hewins (1991) and references therein.

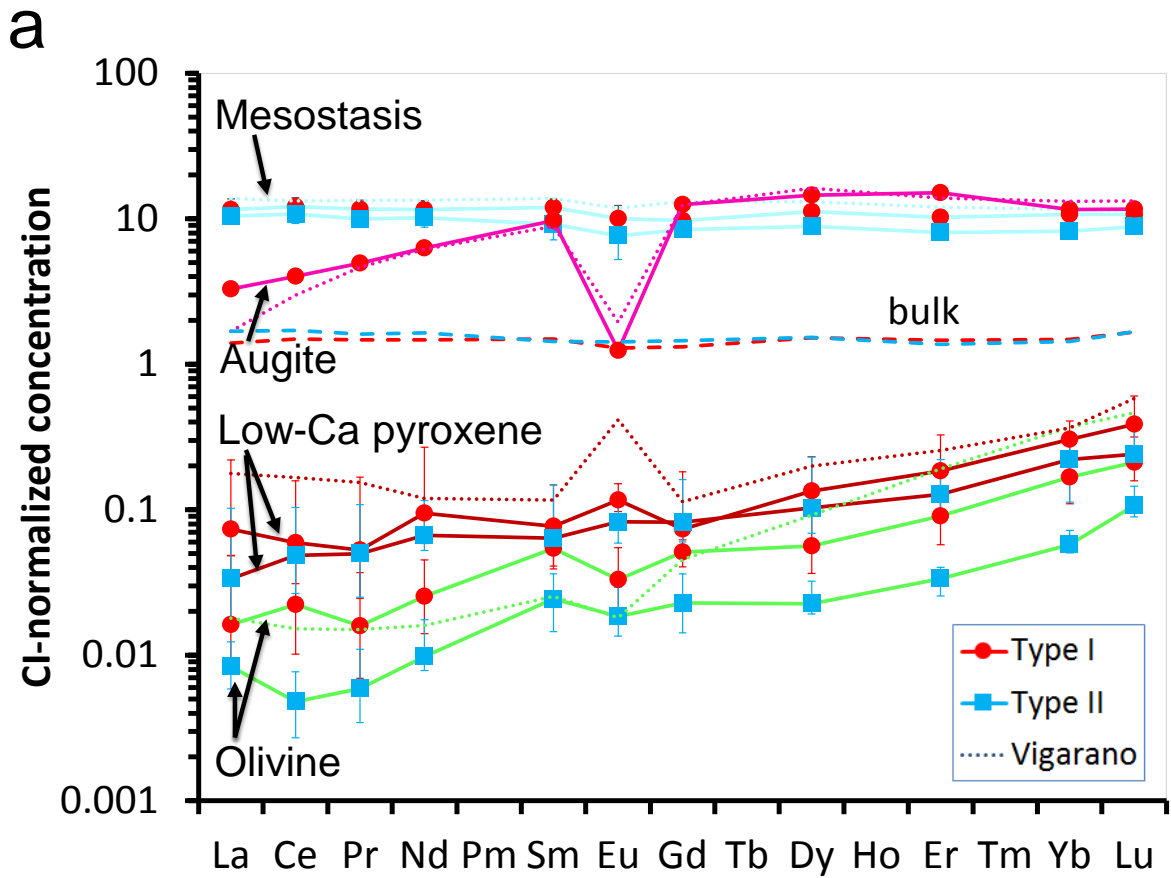

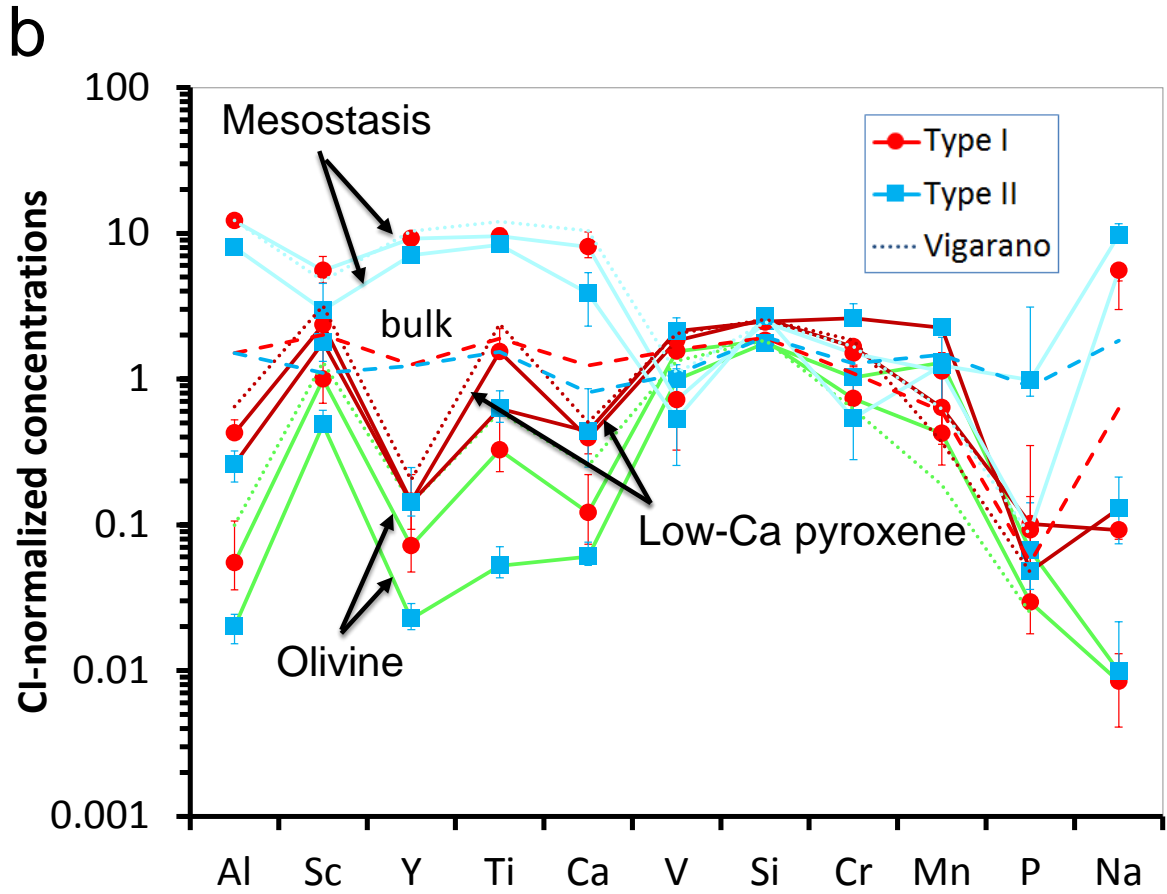

**Figure 2**: (a) CI-normalized rare earth element patterns of different chondrule phases (solid lines) and reconstructed bulk chondrules (dashed lines). (b) CI-normalized average concentration of lithophile elements arranged in order of increasing volatility for olivine, low-Ca pyroxene, mesostasis and reconstructed bulk. Vigarano data (Jacquet et al. 2012) are shown as dotted lines. Error bars extend to 1st and 3rd quartiles.

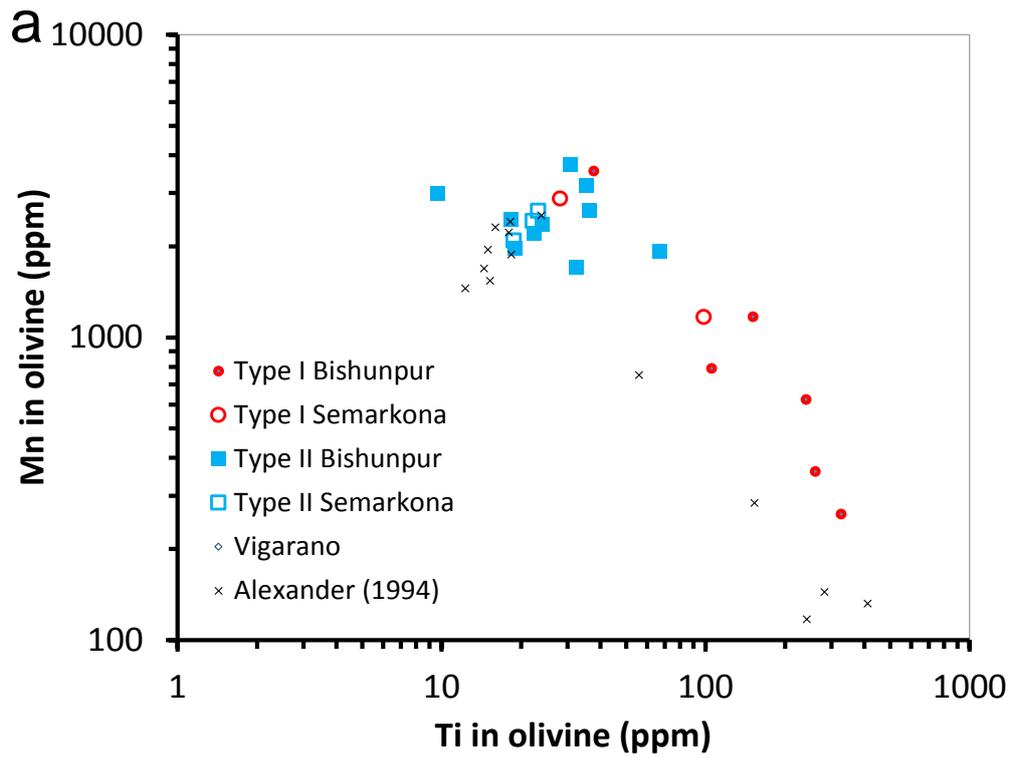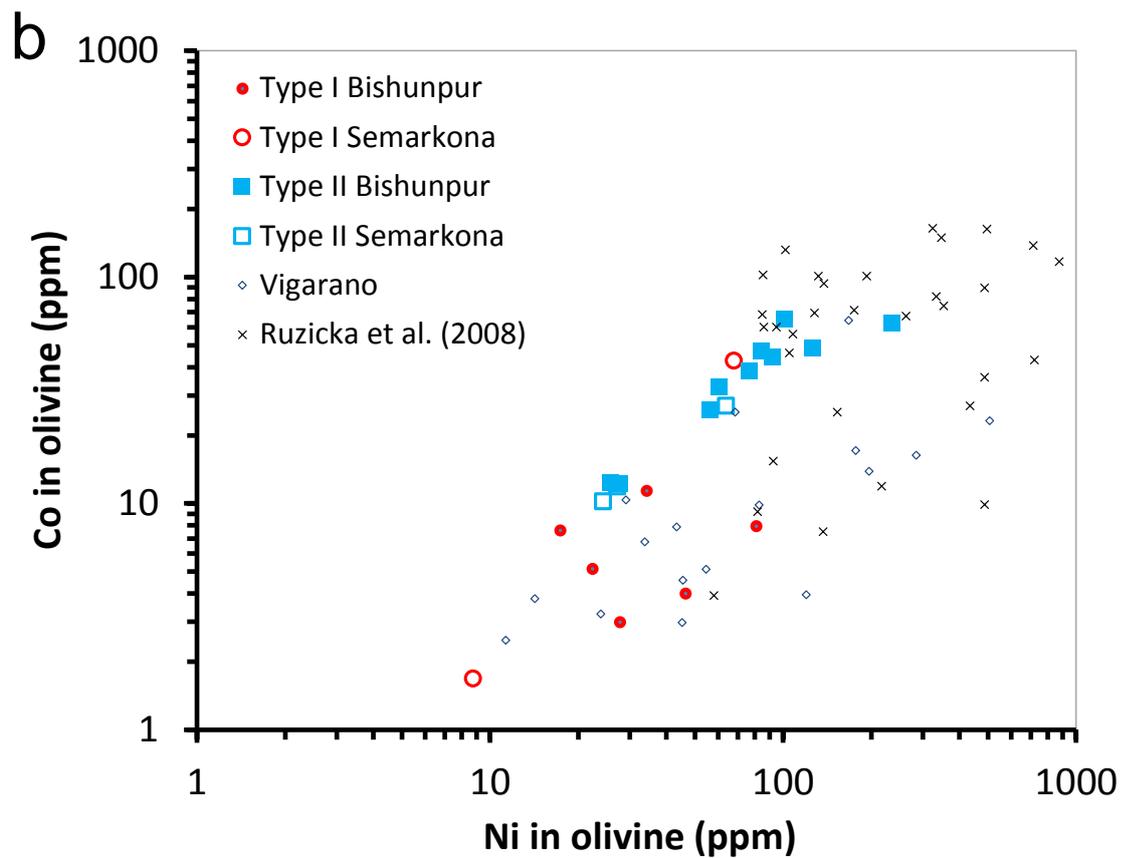

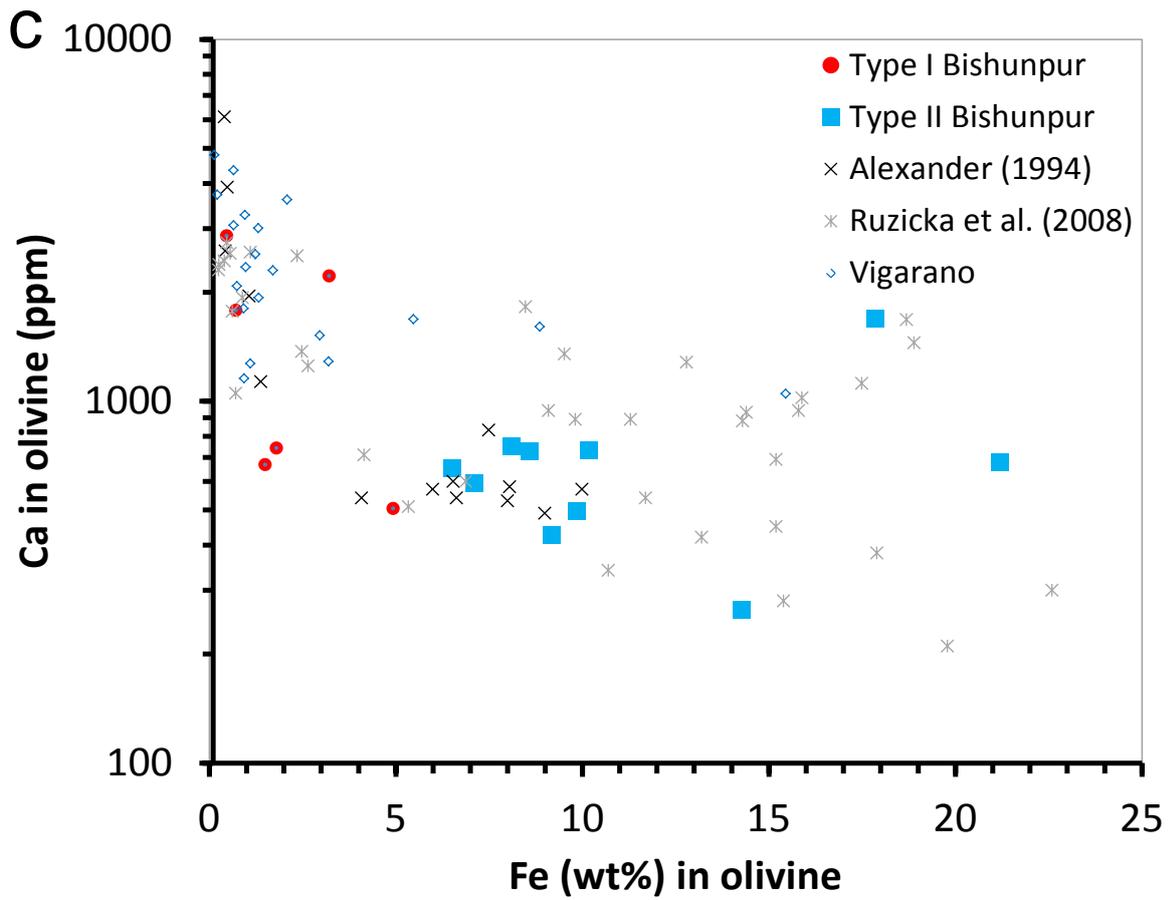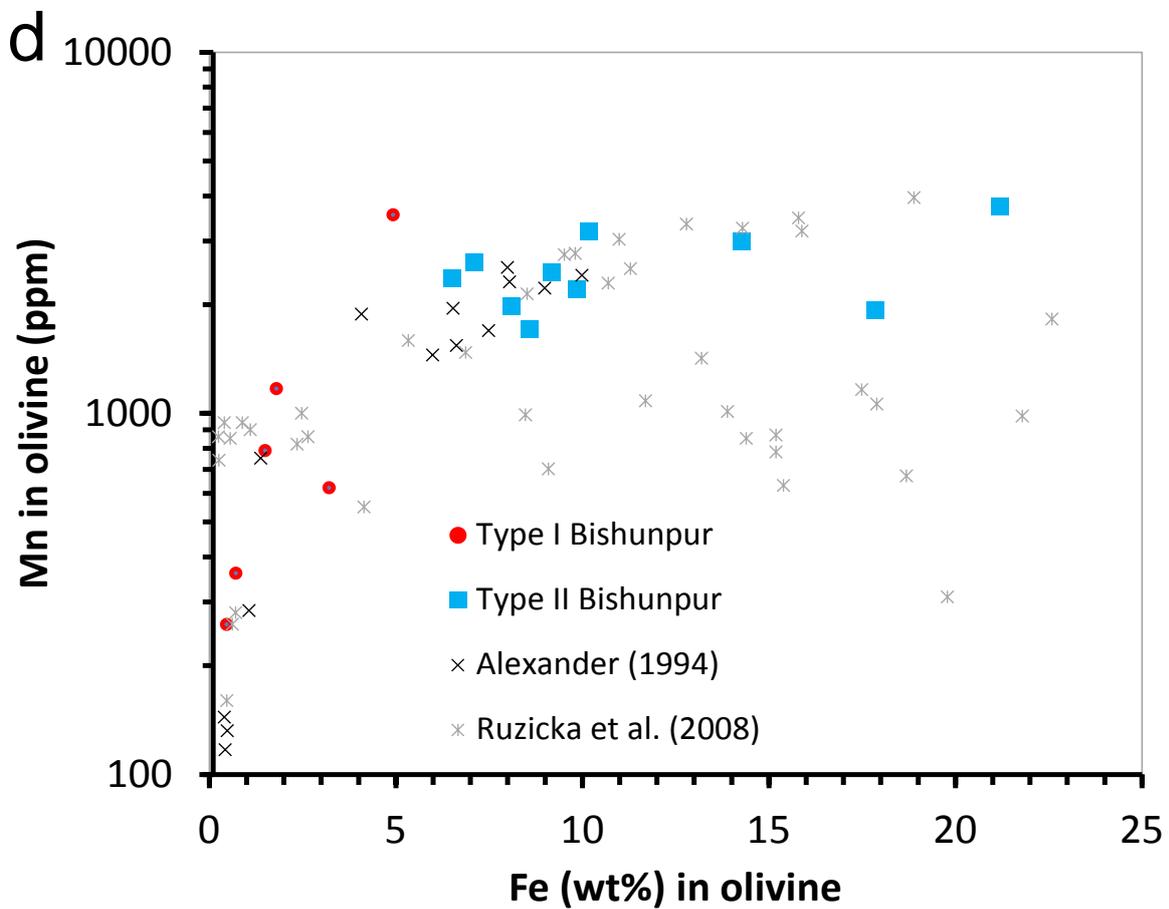

**Figure 3**: Minor element plots in olivine. (a) Mn vs. Ti (b) Co vs Ni (c) Ca vs. Fe (d) Mn vs. Fe. When applicable, comparable data are plotted for Vigarano (Jacquet et al. 2012), or SIMS analyses of chondrule olivine (Alexander 1994; Ruzicka et al. 2008).

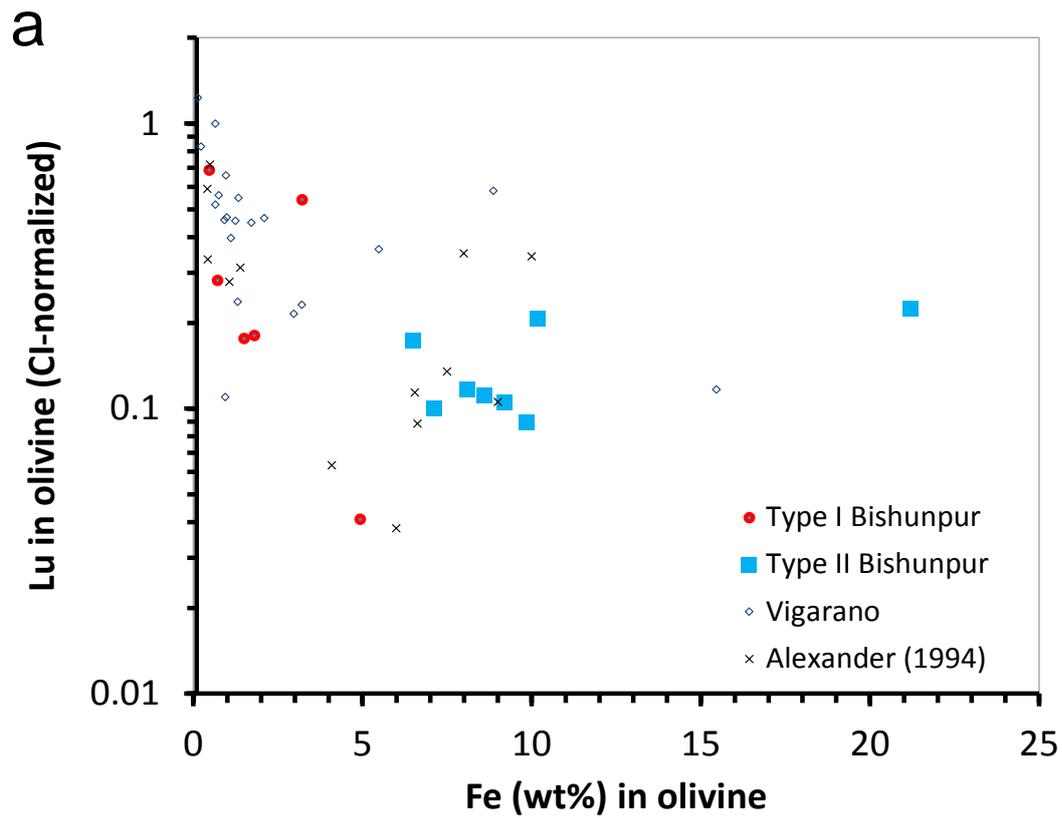

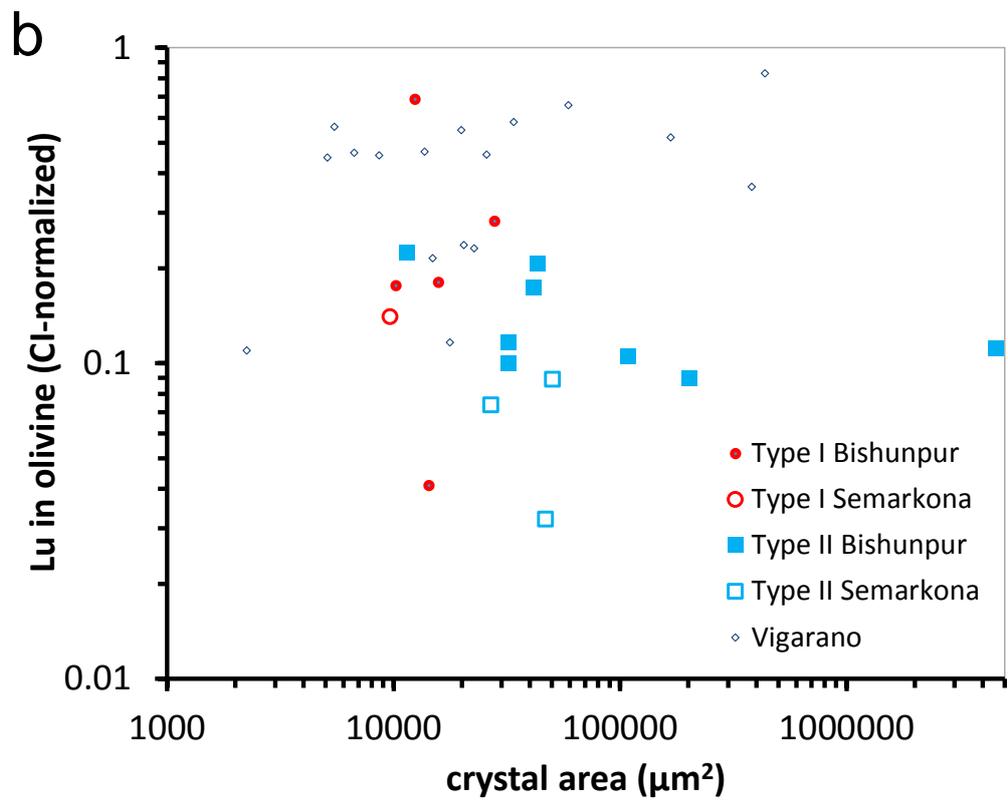

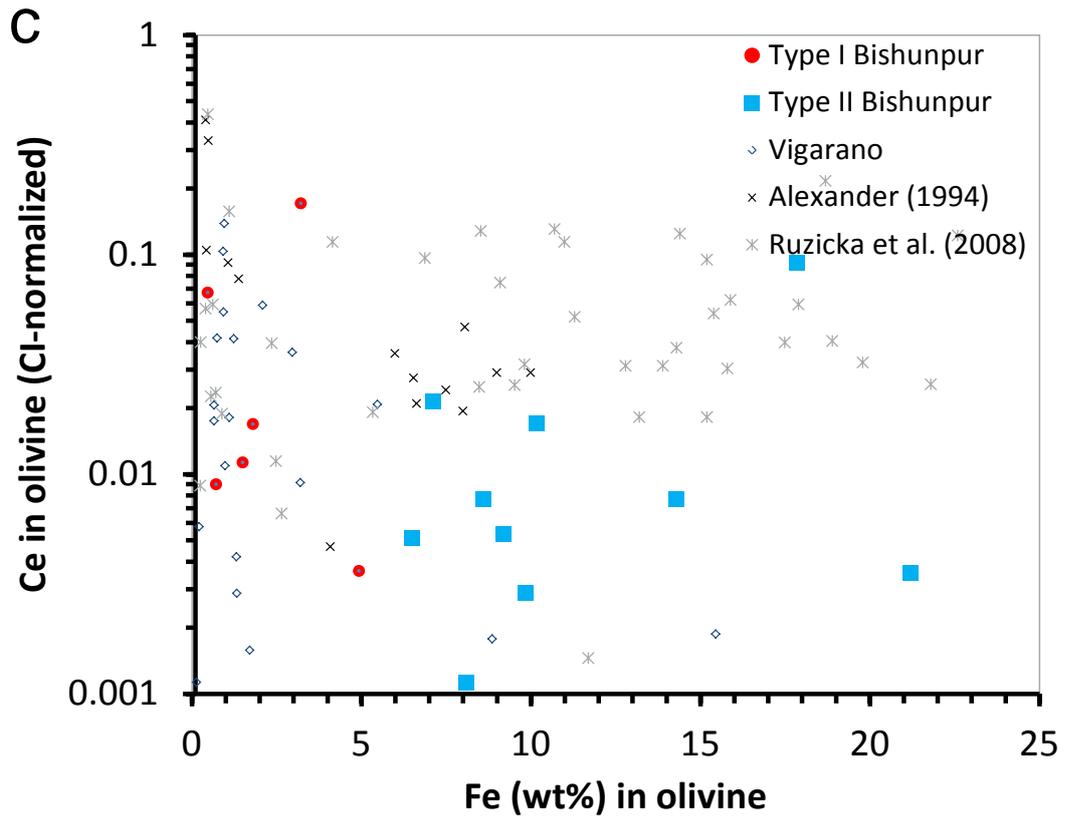

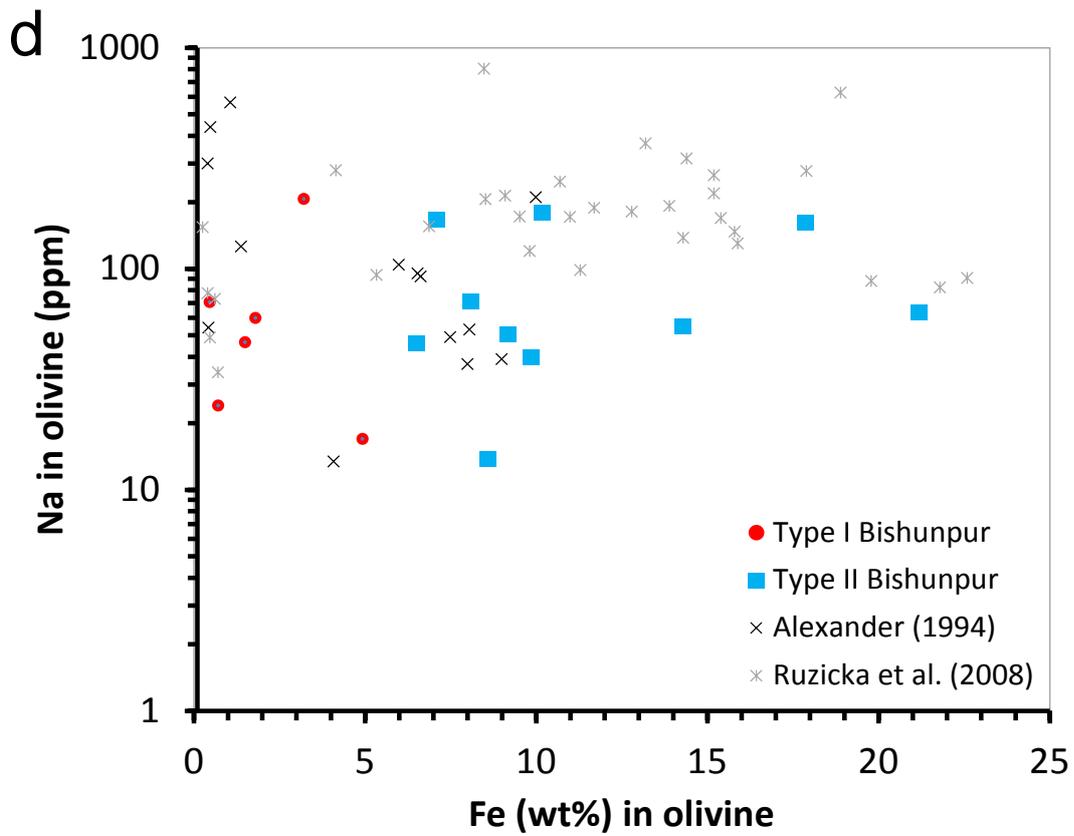

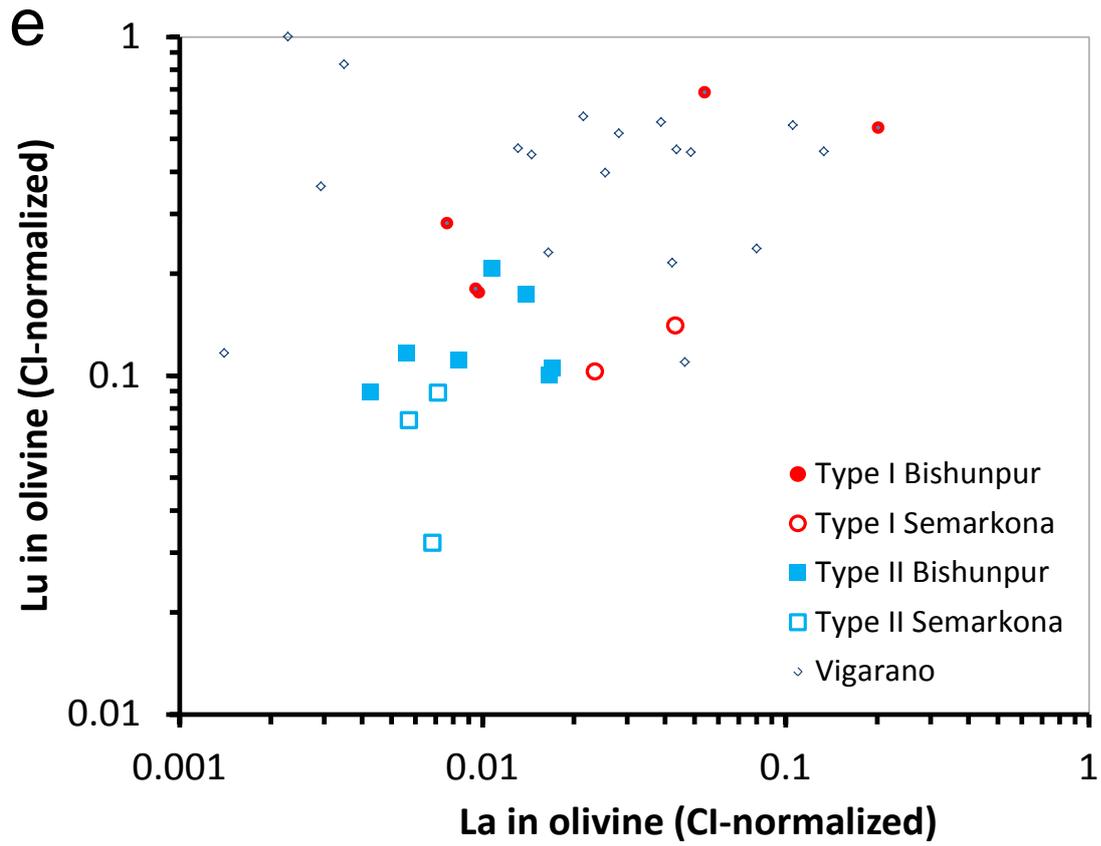
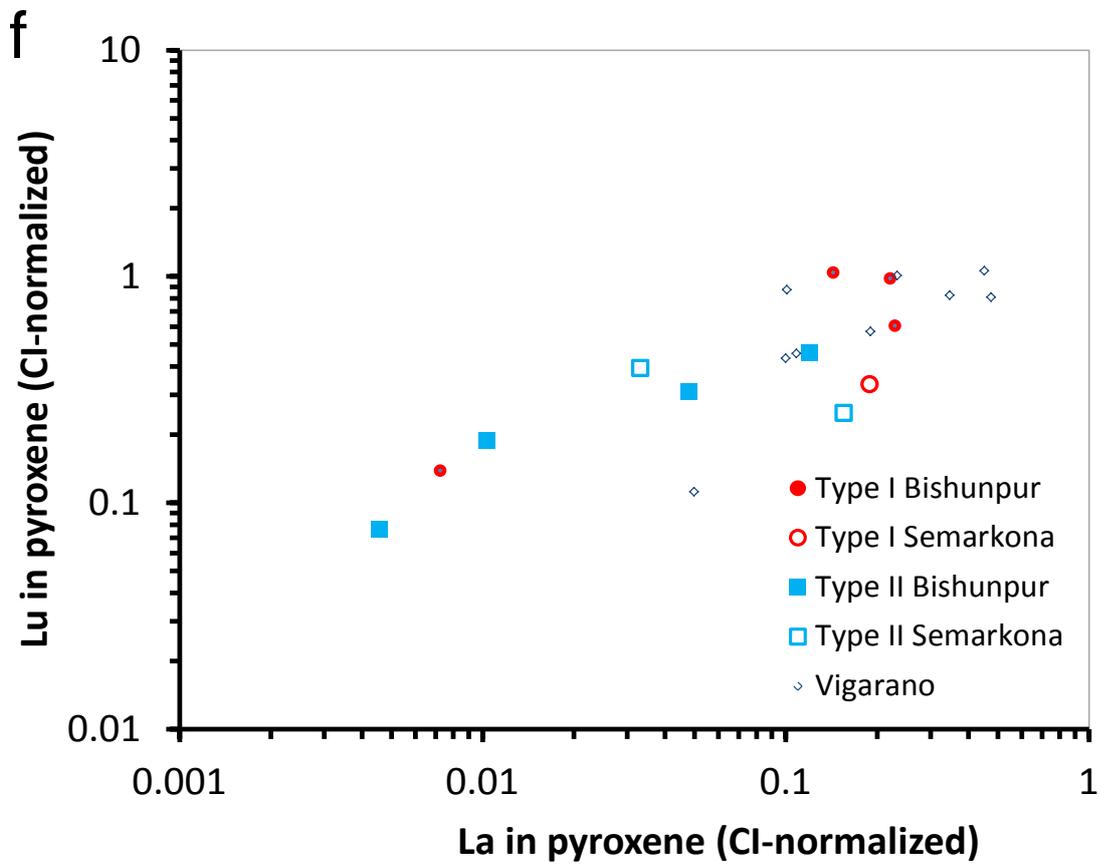

**Figure 4**: Trace element plots. (a) Lutetium vs. Fe in olivine. (b) Lutetium vs. crystal area (averaged over the crystals used to calculate the average composition). (c) Cerium vs. Fe in olivine. (d) Sodium vs. Fe in olivine. (e) Lutetium vs. lanthanum in olivine. (f) Lutetium vs. lanthanum in low-Ca pyroxene. Comparable data are plotted for Vigarano (Jacquet et al. 2012), as well as SIMS data of chondrule olivine by Alexander (1994) and Ruzicka et al. (2008).

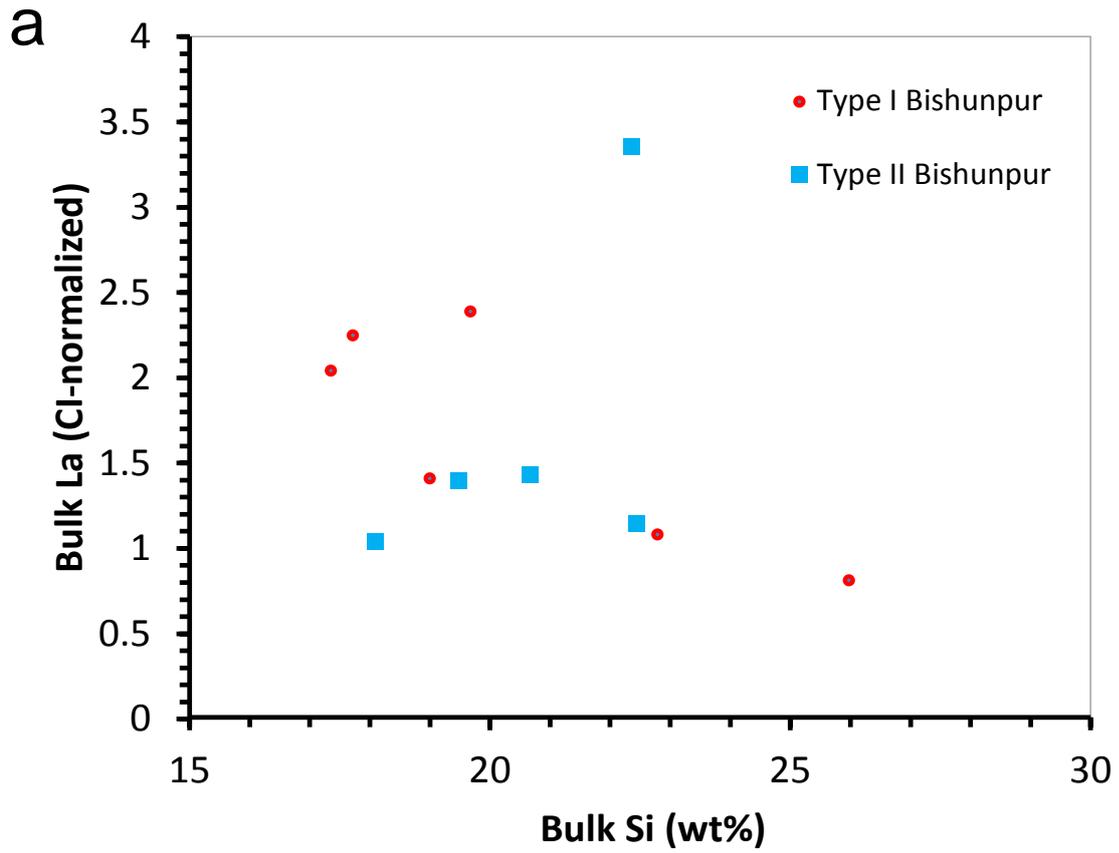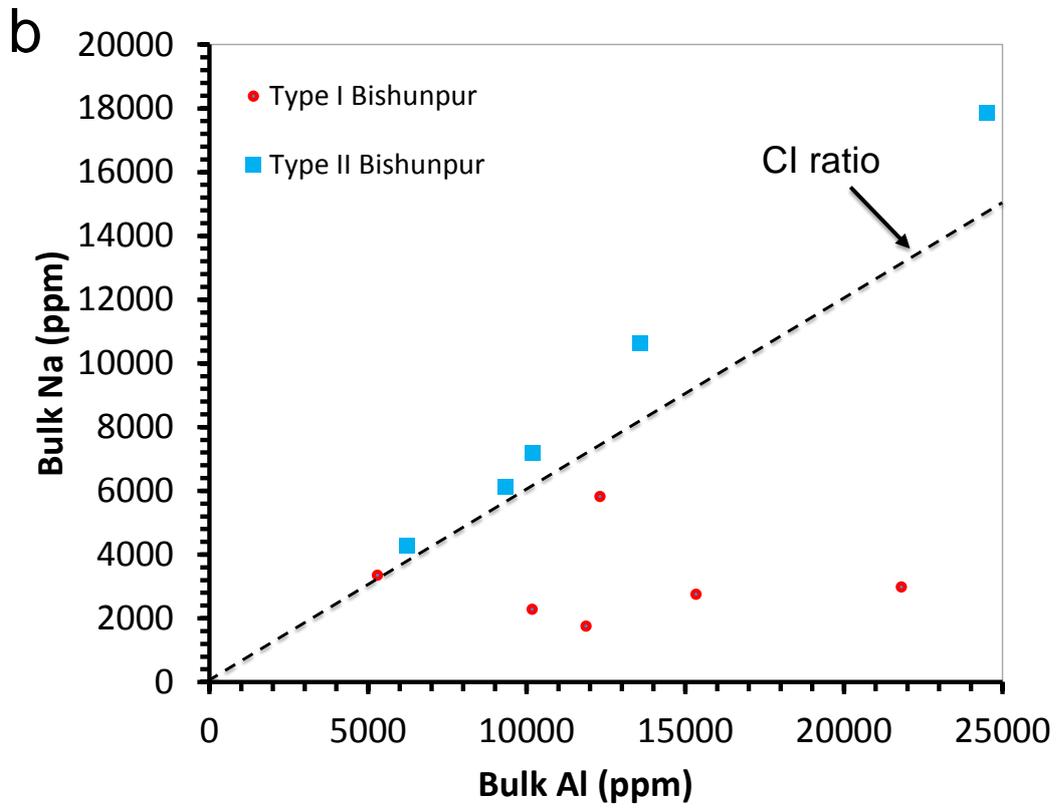

**Figure 5**: Reconstructed bulk chondrule composition plots. (a) Lutetium vs. silicon. (b) Sodium vs. aluminum.

## *4. Discussion*

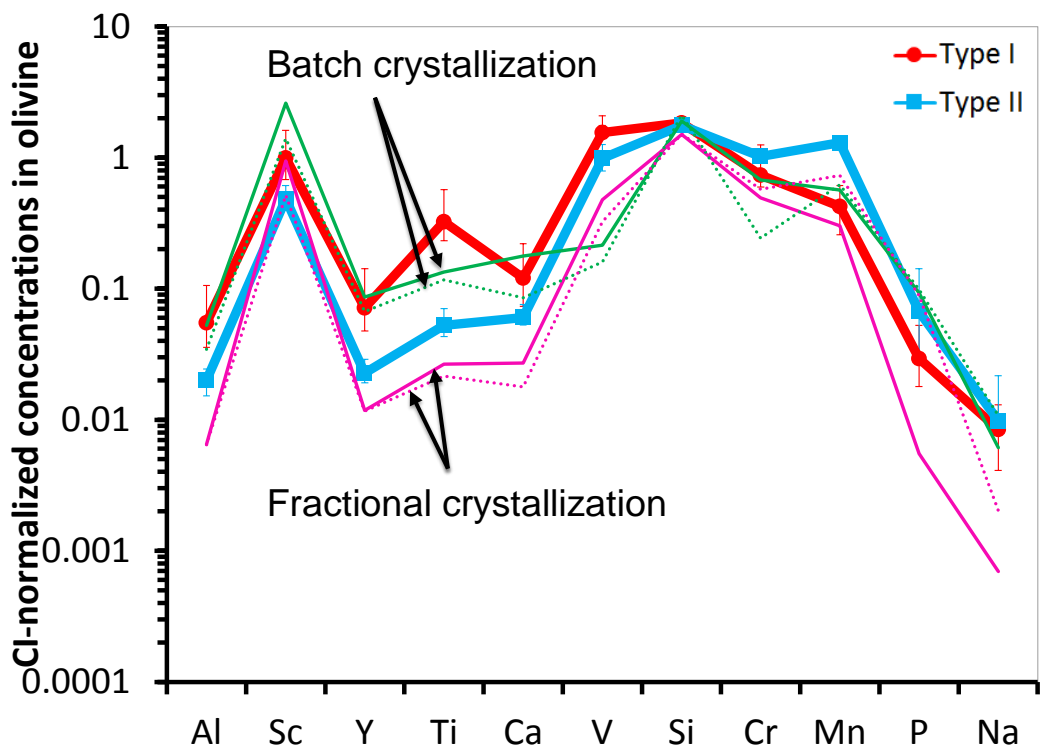

**Figure 6**: Comparison of the observed average olivine compositions for type I and type II chondrules and predictions for batch and fractional crystallization using the observed mesostasis and reconstructed bulk compositions (solid for type I; dotted for type II). Partition coefficients are taken from the PO49 run of Kennedy et al. (1993), with the exceptions of P (Brunet and Chazot 2000) and Na (Mathieu 2009; for forsterite).

### 4.1 Thermal history: type I versus type II chondrules

*4.1.1 Batch vs. fractional crystallization*

In Fig. 6, we compare the average composition of olivine from each type with that calculated from two endmember crystallization models:

1. A *fractional* crystallization model, where the olivine cores (what LA-ICP-MS analyses essentially access) did not change their composition since they solidified and are thus in equilibrium with the *starting* liquid composition, taken to be the average reconstructed bulk from each type.
2. A *batch* crystallization model where, owing to efficient diffusion, the entire olivine crystals maintained equilibrium with the melt now solidified as the chondrule's mesostasis.

We used olivine/melt partition coefficients drawn from experiment "PO49" of Kennedy et al. (1993) conducted at 1525 °C with oxygen fugacity 0.5 log unit below the Iron-Wüstite buffer, with a starting mix of mafic composition, yielding $Fo_{95}$ olivine.

Let us focus first on type I chondrules. As for their carbonaceous chondrite counterparts (Jacquet et al. 2012), it is seen that batch crystallization reproduces the observations better than fractional crystallization, which is off by typically one order of magnitude for refractory lithophile elements. As discussed by Ruzicka et al. (2008) and Jacquet et al. (2012), the anomalies for Ti and V may reflect differing valences of those elements in the conditions of chondrule formation compared to those of the Kennedy et al. (1993) experiments; perhaps inherited from precursors formed under reducing conditions owing to sluggish kinetics (Simon et al. 2011). Ruzicka et al. (2008) proposed that the refractory element enrichment of the liquid in equilibrium with type I chondrule olivine could be understood in a fractional crystallization scenario if chondrules started off refractory and their compositions evolved by condensation during cooling. The one-order-of-magnitude discrepancy between that liquid's calculated composition and the current chondrule composition in refractory elements would however require that typically more than 90 % of the chondrule mass is due to this condensation (Jacquet et al. 2012), at variance with petrographic evidence that chondrule largely formed from the melting of solid precursors (e.g. Hewins et al. 2005). (This does not exclude smaller degrees of gas/solid exchange, see next subsection). We thus believe that the relatively high refractory lithophile element concentrations in type I chondrule olivine is due to equilibration with the mesostasis, which differs from the bulk composition because of the *incompatibility* of these elements rather than their volatility.

If we now turn our attention to type II chondrule olivine, we see in Fig. 6 that the observations are closer to a fractional crystallization scenario, as we had seen for the few type II chondrules we analysed in carbonaceous chondrites (Jacquet et al. 2012). Equilibrium with the initial melt composition would explain the lower refractory lithophile abundances, while higher abundances of volatile elements in olivine reflect the higher bulk abundances of type II chondrules (which are less volatile-depleted than their type I counterparts). A complication ignored by Fig. 6 is that partition coefficients may have been different in type I and type II chondrules. Specifically, experimental olivine/melt partition coefficients of many elements such as REE are anticorrelated with the MgO content of the melt (Colson et al. 1988; Bédard 2005). This presumably would bring theoretical compositions in better agreement with observations ; at any rate, this widens the discrepancy between the liquids in equilibrium with type I and type II chondrule olivine, in the direction indicating batch and fractional crystallization for the two respective categories. This would be consistent with the pronounced minor element zoning for the latter (Jones 1990) compared with weak (or absent) zoning of the former (Jones & Scott 1989).

While we have reasoned above in terms of a batch vs. fractional crystallization alternative (along with all intermediate solutions), a third possibility would be that some olivine crystals are relicts whose unequilibrated composition record a thermal episode anterior to the formation of the host chondrule (e.g. a previous chondrule-forming event). Most recognizable among them are forsteritic grains in type II chondrules and "dusty" olivine grains (i.e. speckled with low-Ni iron grains presumably formed during reduction) in type I chondrules (Jones 1996). In the course of this study, no analysable forsteritic relict was present in the type II chondrules selected and while a few dusty olivine grains have been analysed in type I chondrules, they did not differ from "normal" grains in those chondrules as regards lithophile elements (in agreement with Ruzicka et al. 2008), consistent with extensive equilibration. Nonetheless, Villeneuve (2010) pointed out that originally forsteritic grains may have been completely reequilibrated in type II chondrules by Fe-Mg interdiffusion, leaving open the possibility that type II chondrules may have simply formed by the more or less complete oxidation of Fe-rich type I chondrules, as suggested by the occurrence — especially in carbonaceous chondrites — of entire "grapes" of low-FeO relicts in type II chondrules (see e.g. Fig. 10a of Krot et al. (2007)). But in that case, the REE budget of olivine in type II chondrules would have been essentially either (i) inherited from its type I chondrule "past" or (ii) reset by diffusion from the surrounding melt (or some intermediate solution). (i) is contradicted by the systematic difference in REE budget between type I and type II chondrule olivine, while (ii) would be effectively equivalent to a batch crystallization scenario, inconsistent with the observed REE depletion, as excluded in the previous paragraph. While conversion from type I to type II chondrules certainly occurred, we are thus not inclined to think of type II chondrules as a whole as simply the high-total Fe, oxidized tail of a chondrule population. In fact, the total Fe distribution of chondrules is bimodal (e.g. Fig. 45 of Villeneuve (2010); Wasson et al. 2000), with the type II chondrules constituting the higher-Fe peak, indicative of a qualitative difference between the two. It may incidentally be further objected that type I chondrules from CR chondrites do not tend to be iron-depleted (e.g. Ebel et al. 2008). We thus suggest that simple closed-system oxidation of type I chondrules is the exception rather than the rule for type II chondrule formation (at least for ordinary chondrites). Also, the very preservation of low-FeO relicts in type II chondrules is further evidence that time was insufficient for diffusion throughout olivine crystals to be completed in these chondrules. The absence of high-FeO olivine in type I chondrules (except the rare weakly melted ones like Smk24 (Fig. 1g)) would be consistent with longer thermal processing after which only metal blebs would remain as evidence of the more ferroan composition of the original crystal ("dusty olivine"), although their lower melting temperatures alone might have also led to their frequent destruction.

Yet another alternative to single-stage (batch/fractional) crystallization was developed by Wasson (1996); Wasson and Rubin (2003); Wasson et al. (2014). They proposed that chondrules result from a *series* of small, short-lived heating events (about ~10), possibly (re?)initialized with a larger one, each of them remelting the mesostasis and inducing a few-micron growth of the phenocrysts (as possibly exemplified by oscillatory zoning in some pyroxene grains (Wasson et al. 2014; e.g. chondrule Bi21 in Fig. 1k) although those could simply reflect boundary layer effects (Jones 1996)). In this case, rather than the timescale of one heating event, the trace element systematics would constrain the total time spent at high temperatures (or more precisely, the total time integral of the trace element diffusivity), with type I olivine calling for a longer duration, with either longer or more numerous heating events. Although our above trace element considerations alone would not distinguish between single and multiple events, we comment that the purported different multiple heating events would have to be correlated in some way. If they were not so, the number of heating events undergone by a given particle in chondrites would be a Poissonian of average $N \sim 10$ (and standard deviation $\sqrt{N} \sim 3$). The probability to escape heating altogether would be $e^{-N} = 5 \times 10^{-5}$ for $N = 10$ (and $N e^{-N} = 5 \times 10^{-4}$ for having exactly one heating event), a vanishingly small number to account for unmelted CAIs or even presolar grains which would also have wandered in the nebula (and certainly were among chondrule precursors). Clearly, the multiple heating events postulated by Wasson (1996) would have to be correlated somehow, e.g. because they could occur only (if stochastically) in specific regions of the disk which chondrule precursors happened to cross; so on a global level it would be a matter of language convenience to speak of several events or of one event with nonmonotonic temperature variations. Here, for simplicity, the discussion will be continued in the picture of one heating event, with the understanding that the constraints may be converted in terms of *total* heating/cooling time for multiple events, but we will return to the physical arguments motivating the latter possibility in the course of section 4.2.

*4.1.2 Cooling rates*

The difference between batch and fractional crystallization lies in how the diffusion time $a^2/(2D)$ (with $a$ the crystal radius and $D$ the diffusion coefficient) compares with the cooling timescale in the olivine crystallization interval: if the diffusion time is short compared to the latter, batch crystallization applies, while the converse leads to fractional crystallization, with all intermediate solutions of diffusion-modified fractional crystallization in between. Why is there a difference between type I and type II chondrule olivine in this respect (respectively assigned to the batch and fractional crystallization endmembers)? While olivine crystals in type II chondrules tend to be bigger than their type I counterparts, the entailed increase of the diffusion timescale would be limited : excluding the giant chondrule Bi1, the mean cross-sectional area of analysed type II chondrule olivine crystals is 0.04 mm$^2$ compared to 0.02 mm$^2$ for type I chondrules, with significant overlap. There is also little correlation between REE concentration and crystal size (Fig. 4b), although the smaller type II chondrule crystals tend to be more enriched (presumably because they crystallized later, from a more evolved melt and/or the LA-ICP-MS spots sample a higher fraction of the outer, incompatible element-enriched regions of the crystal). Another possibility would be that diffusion was slower for type II chondrules because of a lower temperature of olivine crystallization. Indeed, taking the representative chondrule bulk composition of Alexander et al. (2008) and judging from the composition-temperature diagrams of Hewins & Radomsky (1990), a systematic difference in liquidus temperature of order 100 K (with significant variations and overlap) is implied. *If* this difference in liquidus temperature could be taken as a proxy for a difference in initial temperature of crystallization (but see Hewins and Radomsky 1990), it would amount to about an order of magnitude difference in diffusion coefficients for an activation energy $E_a \sim 300$ kJ/mol as typical for many elements, e.g. Ca or Fe (Chakraborty 2010). However, this would be essentially counterbalanced by the positive dependence on oxygen fugacity (for a typical 2 order-of-magnitude enhancement of oxygen fugacity inferred for type II chondrules as inferred e.g. by Schrader et al. 2013) and as a typical example, the diffusivity of Ca depends on it to the 1/3.2 power (Chakraborty 2010). The difference in the thermal history of type I and type II chondrules must then lie in the cooling timescale, specifically, type II chondrule olivine records more rapid cooling than their type I counterparts.

Taking $D = 10^{-14}$ m$^2$/s, a typical order of magnitude of diffusion coefficients around 1600 K for many elements (Chakraborty 2010), the dividing timescale would be of order a day. For a "smooth" (e. g. homographic; Dodson 1973) cooling law, with a temperature interval of $RT^2/E_a$, this would translate in a cooling rate of order 10 K/h. In agreement with this, fits of olivine zoning in type II chondrules yield cooling rates generally above this range (e.g. 0.7-2400 K/h (Miyamoto et al. 2009); 700-3600 K/h (Béjina et al. 2009)), while experiments at cooling rates below 10-100 K/h yield no significant zoning in olivine (see Hewins et al. 2005). We note that Huang (1996) also suggested that their "group A" chondrules (roughly equivalent to type I) cooled more slowly than "group B" chondrules (~type II) and maintained equilibrium between mesostasis and phenocrysts. It is recalled that Alexander (1994) had argued against cooling rate as an explanation for the difference between incompatible-enriched and –depleted olivine on the basis of the similar (porphyritic) textures of their host chondrules, but as emphasized by Lofgren (1996), chondrule texture has only a subordinate dependence on cooling rate, as opposed to melting, and should be porphyritic for a wide range of cooling rates (so long chondrules were not heated too long above liquidus (Hewins and Radomsky

1990)). At any rate, we do not think of the rough 10 K/h divide as a strict boundary between type I and type II chondrules, given some overlap in trace element budget (especially above a few mol% fayalite in olivine), and indeed some type I chondrule olivine present some (weak) minor element zoning too (Jones and Scott 1990).

It must be cautioned that above, we have loosely spoken of a diffusion coefficient for all minor and trace elements. In reality, diffusion coefficients certainly depend on the specific elements (Chakraborty 2010), so that they may not be homogenized simultaneously, but at temperatures relevant for olivine crystallization (say 1500-1900 K), these diffusion coefficients are actually comparable within about an order of magnitude (which is one manifestation of the "compensation rule"; e.g. Brady & Cherniak 2010; Béjina et al. 2009) – and it is beyond the scope of this paper to discuss diffusion timescales better than the order-of-magnitude level (which would require dedicated minor element zoning fitting, especially for the type I chondrules hitherto neglected in this regard in the literature). There is, it is true, some uncertainty on whether this extends to REE. Measurements from Spandler et al. (2007) and Spandler & O'Neill (2009) suggest that this is the case, consistent with the equilibration timescale noted in the olivine/melt partitioning experiments of McKay (1986). However, Cherniak (2010) and Remmert et al. (2008) find diffusivities about 3 orders of magnitude lower. The reason of this discrepancy, and whether this may relate to the high REE concentrations used by the latter authors (much higher than natural samples), is currently unknown. If however the latter data were relevant, diffusion timescales numbering in months or years would be implied, which would be conceivable only if olivine grains are relict debris from planetary interiors as envisioned by Libourel and Krot (2007). We argued against this from REE systematics of granoblastic olivine aggregates (GOA; which would be the most pristine samples of these planetary bodies) in our previous study of CV and CR chondrites (Jacquet et al. 2012). We are thus led to prefer the higher values for REE diffusivities for the purpose of this paper.

Alexander (1994), Jones and Layne (1997) and Ruzicka et al. (2008) noted that olivine and low-Ca pyroxene had LREE (and other very incompatible elements) contents much in excess of mineral/melt partitioning predictions. This observation, which we noted extends to carbonaceous chondrite chondrules (Jacquet et al. 2012), is confirmed in this study, if at a lower level than reported by these authors for olivine. Along with the above authors, we ascribe this to kinetic effects, but we have argued against a more quantitative use of the LREE enrichment to calibrate cooling rates in view of nonsystematic variations in experiments and natural samples (Jacquet et al. 2012). In fact, the exact mechanism underlying this "kinetic effect" is still quite uncertain. While the flattening of the REE pattern at the LREE branch (indicative of a convergence of the partition coefficient) suggests entrapment of a cryptic (sub-micron size) melt component (Jacquet et al. 2012b; see also Kennedy et al. 1993), the apparent correlation between LREE and HREE at least for pyroxene (Fig. 4f,g) is unexpected since HREE would be essentially proportionally unaffected by the small inferred levels of contamination. But if another process (e.g. an incompatible element-enriched boundary layer effect as derived by Albarède (2002)) were responsible, one would still need to explain the above convergence of the apparent partition coefficient for very incompatible elements.

While the above has focused on olivine, we note that although slow cooling rates are inferred for its crystallization for type I chondrules, the monoclinic structure of low-Ca pyroxene still seems to require cooling rates in excess of 100-1000 K/h around 1300 K (Jones and Scott 1990), and hence a nonlinear cooling history, as advocated for carbonaceous chondrite chondrules by Jacquet et al. (2012, 2013). The cooling rate inferred for type I chondrule enstatite is interestingly in the range inferred for the olivine of type II chondrules, which Ferraris et al. (2001) verified applied to their pyroxene as well, from whose nanostructure they inferred cooling rates of 50-3000 K/h around 1500

K, before slowing down at <10 K/h below 1000 K. That is, the thermal history of type II chondrule could differ from that of type I chondrules simply in lacking the slower cooling stage (or having only a short one; e.g. Villeneuve (2010)). The rare type I chondrules with ferroan, non-dusty relicts like Smk24 (Fig. 1j) might also have lacked the slow cooling phase. Perhaps, then, the higher FeO content of olivine in pyroxene-rich type I chondrules relative to type IA chondrules (e.g. chondrule Bi9; Brearley and Jones 1998) may also indicate somewhat higher oxygen fugacities at this stage as a further analogy.

## 4.2 Open system behavior

It is widely recognized that chondrules did not behave as strictly closed system but interacted with ambient gas, by evaporation, recondensation and isotopic exchange (Libourel et al. (2006), Hewins & Zanda (2012), Marrocchi & Libourel 2013)). Like Ruzicka et al. (2008), we note anticorrelations between refractory and moderately elements, e.g. Ti and Mn (Fig. 3a). As such, these could reflect either vapour fractionation during chondrule formation or mixing between variously refractory precursor grains. Similarly to Huang et al. (1996), we suggest that the systematic difference between type I and type II chondrules, with the former more volatile-depleted than the latter, which have different thermal histories (as inferred above), indicate that part of the inferred vapour fractionation took place during chondrule formation, with more volatile element loss in type I chondrules. A precursor effect due to different nebular reservoirs seems further unlikely as type I as well as type II chondrules are different in the different chondrite clans (Jones 2012) so that many distinctive subreservoirs would be needed whereas the refractory lithophile element fractionation as recorded by bulk chondrites (Larimer & Wasson 1988) appears to have been a global (disc-wide) process. Plausible explanations for the larger depletion of volatile elements for type I chondrules may be the longer timescales during they were processed at high temperatures, as we inferred, the lower oxygen fugacities which destabilize solid oxides (by Le Chatelier's law), or lower concentrations of solids for which saturation by the evaporated species would be possible only for larger evaporated fractions (e.g. Alexander et al. 2008). The overall depletion in Fe of type I chondrules relative to type II chondrules (which are themselves depleted relative to CI chondritic abundances) may also reflect the volatility of Fe (which Mullane et al. (2005) showed to be isotopically heavier for type I chondrules). Nonetheless, loss of iron as metal (which would indeed affect mostly type I chondrules), from surface energy effects as calculated by Uesugi et al. (2008), or owing to inertial acceleration as invoked for cosmic spherules (Genge & Grady 1998), is also conceivable (and petrographically supported in CR chondrites (Campbell et al. 2005)).

In the remainder of the section, we discuss evidence from two specific elements, silicon and sodium.

*4.2.1 Silicon*

The fact that the Si-richest (pyroxene-rich) type I chondrules are poorest in REE (or other refractory elements; Fig. 5a) is consistent with a dilution of the latter by the former. This is what Libourel et al. (2006) inferred based on similar anticorrelations in chondrule mesostases, and the general concentration of pyroxene (whose crystallization would be promoted by silica addition from the outside) at the chondrule margins. In our earlier work on carbonaceous chondrites (Jacquet et al. 2012), we had supported this idea from an anticorrelation between enstatite/olivine REE partition coefficient and enstatite mode, which our few enstatite data do not allow us to confirm here, but this might be too indirect a proxy anyway compared to bulk composition (although it would explain the apparent discrepancy between the data of Alexander (1994) and the lower values by Jones and Layne (1997) as the latter authors studied more pyroxene-rich chondrules). Libourel et al. (2006) proposed that this silica addition was due to reaction of the melt with SiO gas. The frequent observation of small pyroxene-rich microchondrules either adhering to type I chondrules or trapped in their Fe-rich fine-grained rim (Fig. 1b-e) suggests an additional pathway, namely by accretion onto molten chondrules of droplets enriched in silica, possibly via the same reaction envisioned by Libourel et al. (2006) but sped up by their enhanced surface-to-volume ratio. These droplets might be ultimately derived from the fragmentation (during collision or ram pressure stripping) of normal-size chondrules as proposed by Dobrica et al. (2014) and Bigolski et al. (2013), although the protuberances presently observed on chondrule margins may simply be microchondrule adhesions (sometimes supported in that by sulphide linings between them and the host, e.g. Fig. 1b,c) rather than droplets instantaneously caught in the act of separation. As to microchondrules, this seems to us a more natural scenario than formation by limited remelting of the primary chondrule and immediate trapping in dust as proposed by Krot et al. (1997). Although the compositional similarity between microchondrules and chondrule edges as well as the occurrence of protuberances on those edges do not discriminate between these two scenarios (which are in a sense time-reversed mirrors of each other), our scenario would avoid having to postulate another thermal episode and very rapid formation of the fine-grained rim.

Accretion of droplets on chondrules was already invoked by us for CR chondrite type I chondrules in our study of their metal trace element geochemistry (Jacquet et al. 2013), and may thus be a general phenomenon (although the FeO-rich fine-grained rims (with trapped microchondrules) are apparently uncommon in CR chondrites, but one may be seen around chondrule "ch4" of EET 92042 (CR2) in Fig. 2 of Connolly et al. (2001)). Hence, such accretion at the end of the chondrule-forming event might explain the igneous rims around ordinary chondrite chondrules as well, which may form a petrographic continuum with the occurrence of protuberances. This might be the simplest way to account for the similar FeO contents of the igneous rim silicates relative to their host chondrules' (Krot & Wasson 1995), rather than invoking separate accretion and re-heating events (e.g. Rubin and Wasson 1987) under the same redox conditions — which appeared noncoincidental only in the old, monotonically cooling "hot solar nebula" picture (Krot & Wasson 1995). Tying hereby igneous rim formation to chondrule formation in a single event would also help explain why only a very small minority of CAIs, which should have been exposed to those in the disk, display these igneous rims (Rubin 1984). At any rate, our suggestion does not, however, exclude that *some* igneous rims formed by limited heating (e.g. in the vicinity of a chondrule-forming region) of fine-grained rims accreted after a first cooling, especially in order to account for igneous rims around fragments or instances of multiple layering (Rubin 1984).

*4.2.2 Sodium*

Sodium is a moderately volatile element of especial interest, for it should have been lost from chondrules within ~10 s at high temperatures under "canonical" conditions (i.e. solar abundances; total pressure < 100 Pa; e.g. Fedkin and Grossman 2013). Yet chondrules still contain substantial amounts of sodium (e.g. Fig. 2b), near chondritic values for type II chondrules.

What is more, type II chondrules show a positive Na/Al correlation around an atomic ratio of 0.8 (Fig. 5b; Hewins 1991). Such a correlation between elements of nominally so different volatilities is difficult to account for by vapour-melt interaction. That would hold even upon external buffering of Na: indeed Mathieu et al. (2011) and Mathieu (2009) experimentally found that the activity coefficient of $Na_2O$ in silicate liquids, to which the solubility is inversely proportional, increased with the optical basicity. As $Al_2O_3$ has a relatively high optical basicity (0.6) and is generally inversely correlated with $SiO_2$ in chondrule mesostases (e.g. Libourel et al. 2006; this study), which has a low one (0.48), higher Al contents would lead to high optical basicities and therefore low Na solubilities. The correlation between Na and Al may then have to be attributed to a precursor effect. It could e.g. simply reflect igneous fractionation, Na and Al being both incompatible, with chondrule mesostasis being among the precursors in varying amounts and varying fractionation degrees in a recycling process (e.g. Alexander 1994), provided the cycle avoided the Na loss experienced by type I chondrules. It could also reflect an albitic component (e.g. Hewins 1991), a plausible source of which might be pre-existing metamorphosed chondrites (or primitive achondrites) which contain of the order of 10 vol% oligoclase feldspar (Brearley and Jones 1998; Mittlefehldt et al. 1998).

It would be then required that type II chondrules remained closed systems for Na. This is supported by the appreciable Na concentrations measured in chondrule olivine (32 ppm (excluding the BO chondrule Bi3) and 49 ppm on average, respectively), which are roughly consistent with partitioning with mesostasis and bulk compositions, respectively (Fig. 6; Alexander et al. 2008). However, Hewins et al. (2012) argued that type II chondrules in Semarkona lost and regained a significant fraction of Na after the onset of olivine crystallization. In particular, they noted that the olivine core/bulk partition coefficient for Na was higher than the olivine rim/mesostasis one in type IIA chondrules, the opposite of what they would have expected for a closed system given a positive correlation of experimental partition coefficients with fayalite contents (Kropf & Pack 2009; Mathieu 2009). However, this positive correlation is only a partial corollary of the positive dependence of the partition coefficient (directly proportional to the activity coefficient of $Na_2O$ in the melt) on optical basicity, given that FeO has a higher optical basicity than MgO (1 vs. 0.78). It does not take into account the fact that the initial liquid (the bulk chondrule) is poorer in silica than the late liquid (after olivine extraction) represented by the mesostasis, yielding a *higher* optical basicity originally. Indeed, from our type II chondrule compositions, we calculated a range of optical basicities of 0.60-0.65 for the bulk and 0.52-0.57 for mesostases, so from Mathieu (2009), the original partition coefficient may have been easily a factor of 3 or so higher than the final one, as observed. We thus conclude that most type II chondrules indeed behaved as closed systems.

Then, how are we to avoid Na loss in the first place? This conundrum is one of the main motivations behind Rubin and Wasson (2003)'s preference for short heating events (see end of section 4.1.1). However, while *one* such event may lead to limited Na loss, what matters is the *cumulative* loss resulting from all such heating events undergone by a given chondrule. Certainly, from the interpretation developed in section 4.1, the total timescale, at least for type I chondrules, would be orders of magnitude in excess of the Na loss timescale. Independently of trace element systematics, a lower limit for the duration of each high-temperature event would be the timescale of pure radiative cooling, which is of order a second (e.g. Wasson 1996). This is most conservative as it ignores the possibility that the ambient gas and radiation field may have sustained high temperatures for longer durations. But if multiplied by 10 — the order of magnitude of the number of events favoured by Rubin and Wasson (2003) —, this already reach the timescale for total Na loss calculated by Fedkin and Grossman (2013). Another lower limit might be obtained from the extent of crystal growth; for instance olivine growth rates measured by Jambon et al. (1992) never exceed 0.6 μm/s, suggesting an order of magnitude longer minimum total time. Thus, although the form of the time-temperature curve is open to debate, it is doubtful whether a radical shortening in one-time heating timescale can resolve the problem.

Yet, if so, the Na conundrum remains. Whether one invokes extremely high concentrations of solids (Alexander et al. 2008), coupled with extremely high total pressures (Fedkin and Grossman 2013; see also Cuzzi and Alexander 2006), or external Na buffering (Morris et al. 2012), there seems to be no way around the conclusion that specifically the *gas* component was far from canonical during chondrule formation (independently of devolatilization from chondrules). This might point to environments in the vicinity of planetesimals or protoplanets, whether gaseous or terrestrial (with outgassed atmosphere). Impact melting (Symes et al. 1998; Fedkin and Grossman 2013) or molten planetesimal collision scenarios (Asphaug et al. 2011) which would satisfy this would have difficulties in accounting for the apparently nebular character of chondrule precursors (Jacquet 2014a). Possible scenarios include (but are not limited to) bow shock by planetary embryos with outgassed Na-rich atmosphere (Morris et al. 2012) or gas dynamic heating in a planetary atmosphere (Podolak et al. 1993).

4.3 Compared environments of formation of type I and type II chondrules

In this section, we would like to sketch the emerging picture on the differences between the environments of formation of type I and type II chondrules, and what these differences may indicate on the chondrule-forming process(es) in general.

Could type I and type II chondrules have formed in the same events, with their oxygen fugacities controlled by their composition (e.g. Connolly et al. 1994)? This appears incompatible with the different thermal histories we have inferred for them, the lack of evidence of silica addition during cooling in type II chondrules compared to type I chondrules (with their frequent pyroxene-rich margins), the differing nature of rimming material (ferroan fine-grained around type I, sulphides around type II chondrules). The correlation between FeO content of silicates of siblings in compound chondrules (Wasson et al. 1995) also suggests an externally buffered oxygen fugacity, while the presence of some trivalent Ti in type II chondrules suggest relatively reduced precursors even for them (Simon et al. 2011). As to compound chondrules, it is also notable that type II chondrules have a higher compound fraction than their type I counterparts: indeed the compound chondrule olivine analyses compiled in table 5 of Wasson et al. (1995), show, if we restrict attention to LL chondrites, 42 analyses corresponding to type II chondrules while only 5 relate to type I chondrule olivine, hence a ratio of about 8, although the global type II/type I ratio does not exceed 2-3 (Zanda et al. 2006). Possibly the less round shapes of type II chondrules relative to their type I counterparts (Scott and Taylor 1983) may also indicate that many more of them are unrecognized compound chondrules with blurred boundaries (see Akaki & Nakamura 2005; Alexander and Ebel 2012). This suggests that both type of chondrules recorded different collision rates, likely due to higher solid concentrations in type II chondrule formation regions (especially if timescale of formation of the latter was shorter as we infer). Wasson et al. (1995), it is true, questioned the usual interpretation of compound chondrules as the outcomes of chondrule-chondrule collisions (Gooding and Keil 1981; Connolly et al. 1994) and interpreted for example the "independent" ones by the flash-melting (in the Wasson (1996) picture) of fine-grained material adhering to the primary chondrule. It would be however difficult to understand the prevalence of very asymmetric rims postulated for the "consorting" compounds in contradistinction to the "enveloping" variety.

This all suggests distinct formation environments for both chondrule types. A conflicting piece of evidence, though, may be the existence of "cluster chondrites", i.e. chondrites with deformed and mutually indented chondrules (like in Semarkona), which Metzler (2012) interpreted as the result of very rapid accretion with the heat inherited from the chondrule-forming event. As these "cluster chondrites" consist of intimate mixtures of type I and type II chondrules, this would imply that both types could form at the same time in the same region. However, we believe that accretion directly after chondrule formation is not necessarily required for cluster chondrites. Indeed, they exhibit presolar grains (Metzler 2013) and rare CAIs with canonical $^{26}Al/^{27}Al$ initial ratio (e.g. Bischoff & Keil

1984); Russell et al. 1996)) which are not expected to survive chondrule-forming events. Moreover, chondrules analysed in Semarkona show a wide range of Al-Mg ages (Villeneuve et al. 2009; but see Alexander and Ebel 2012). It is thus conceivable that cluster chondrites represent well-mixed chondrite components – in fact, they essentially present the same proportions of chondrule textural types as other non-cluster chondrites (Metzler 2012) — from various sources which upon or after accretion were subject to a constraint that led to their deformation (presumably accompanied by some reheating, but not to the extent of melting the interchondrule matrix), although the origin of that constraint (shock?) is still the subject of on-going investigation (Metzler 2012). We note that mutually indented chondrules, rims and refractory inclusions, with no consistent axis of strain, have been reported for the Northwest Africa 3118 CV3 chondrite (Barlet et al. 2012). We thus conclude that type I and type II chondrules formed in different environments, although these environments may have been contiguous at large scales.

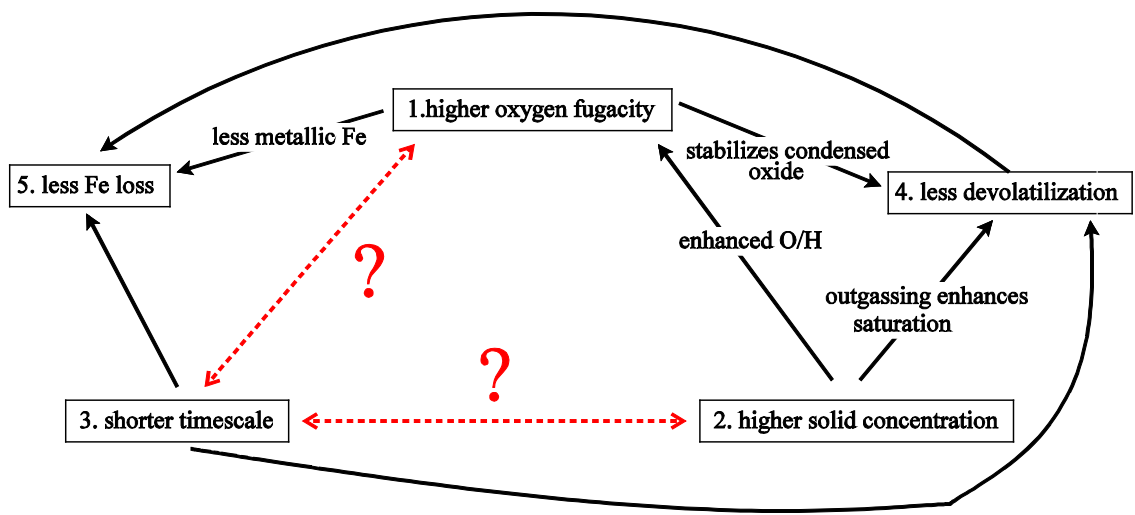

**Figure 7**: Causal diagram of the differences between the formation conditions of type II chondrules relative to type I chondrules (numbers are included for ease of reference in the text). Arrows (solid lines) indicate possible causal links discussed in the text. A link between shorter timescale and higher solid concentrations and/or higher oxygen fugacities (shown in dashed red lines with question marks) has yet to be accounted for—a constraint on chondrule-forming models.

What are the main differences between these environments of formation? To summarize the preceding discussion, we have seen that compared to type I chondrules, type II chondrules likely formed:

(1) under higher oxygen fugacity

(2) with higher solid concentrations

(3) in shorter timescales

(4) with less devolatilization

(5) with less iron loss.

A chondrule-forming mechanism should thus ideally be able to explain all these differences in terms of one variable if the conjunction of these properties is not to be attributed to coincidence — *provided*, what is far from certain, that type I and type II chondrules formed by comparable mechanisms. From a cosmochemical perspective, we have already suggested some causal explanations as first relatively model-independent steps toward such a unification: (1) may be explained by (2) (because of the oxygen-rich nature of dust, with or without water ice), while (4) may be due to (1), (2), and/or (3) and (5) may be ascribed to either (1), (2) or (3) (see "Open system behaviour"). This is summarized graphically in the "causal diagram" in Fig. 7. But fundamentally, one would still need to specifically account for a link (in one direction or the other) between oxygen fugacity (or solid concentration—either absolute or ratioed to hydrogen), and the chondrule formation time. This is thus a physical constraint for chondrule-forming scenarios.

We note as an example that the shock wave model of Desch & Connolly (2002) specifically predicted a positive correlation between cooling rate and chondrule density, which would thus satisfy this contraint. (It may be commented that this prediction was originally emphasized with respect to nonporphyritic chondrules, because of their higher compound chondrule frequency (Gooding and Keil 1981), as they can be reproduced experimentally with rapid cooling rates, but chondrule texture is more dependent on melting than cooling (Lofgren 1996), and the frequency of compound nonporphyritic chondrules may have to do with their intrinsic rheology (Ciesla et al. 2004; Jacquet 2014a); it would however be interesting to investigate to what extent the bias of compound chondrules in favour of nonporphyritic textures relates to that in favour of type II compositions, which typically have lower liquidus temperatures). A similar prediction was recently made in the framework of the impact jetting model by Johnson et al. (2015). Another possibility for relating timescale and concentration might be to invoke the thermal inertia of solids compared to the gas: the energy per unit mass required to melt a solid of volatile-free CI chondritic composition is 1.6 MJ/kg (Sanders and Taylor 2012) — which may be significantly increased by the presence of water — while that required to raise the temperature of hydrogen gas by ~1500 K is of order 10 MJ/kg. For dust/gas ratio above a few hundred times solar, as inferred for type II chondrule formation (e.g. Schrader et al. 2013), the thermal inertia of solids would dominate, and may have thus reduced the time-integrated efficiency of an energetic "pulse". These are only three ideas and other explanations may be conceivable, but discussing them in any comprehensive manner is certainly beyond the scope of this paper, although this question should be addressed by future chondrule formation models.

### 4.4 Comparison to carbonaceous chondrites

Overall, results from this study on ordinary chondrites are quite similar to those we reported for carbonaceous chondrites (Jacquet et al. 2012), with in particular the same difference between type I and type II chondrules. If we restrict attention to type I chondrules which are well-represented in both carbonaceous and ordinary chondrites, we may note that their silicates are systematically poorer in refractory lithophile elements and enriched in moderately volatile ones (Fig. 2). This presumably only reflects the enrichment in refractory elements of carbonaceous chondrites compared to ordinary chondrites (see e.g. Krot et al. 2003)); i.e. a precursor effect. This is consistent with the idea of "local" chondrule production (e.g. Jones 2012), subsequent to refractory/volatile element fractionation among chondrite reservoirs.

It would seem that the primary difference between the two classes lies in the relative abundance of type I and type II chondrule producing events, the latter being more prevalent in the source reservoirs of ordinary chondrites, which might e.g. have been related to their position relative to the snow line (Ruzicka 2012). This is likely reflected as well in the widespread occurrence of forsteritic clusters in type II chondrules in carbonaceous chondrites whereas forsteritic relict grains are only encountered in 10 % of type II chondrules in ordinary chondrites (Jones 1996). Not only seem shorter events—as we interpret applied to type II chondrules—more prevalent in ordinary chondrite reservoirs, processing of the type I chondrules themselves may have been less long-lived on average than for their carbonaceous chondrites counterparts. Indeed ordinary chondrites are generally devoid of GOAs (which Whattam et al. (2008) approximately reproduced after days of thermal processing) and of plagioclase-bearing type I chondrules (unlike those in CO chondrites, which Wick & Jones (2012) could only reproduce at cooling rate ≤ 1 K/h at 1300 K). Also, ordinary chondrite olivine exhibit a lack of very low (near-equilibrium) $(Ce/Yb)_N$ ratios (always above 0.02), contrasted with values down to 0.002 which we reported for carbonaceous chondrites, suggesting more kinetic effects there (see section 4.1.2).

A further significant difference lies in the fact that, for carbonaceous chondrites, type II chondrules are systematically depleted in oxygen-16 relative to type I chondrules (e.g. Wasson et al. 2000; Rudraswami et al. 2011; Ushikubo et al. 2013; Tenner et al 2013; Schrader et al. 2013), whereas such a trend largely disappears for ordinary chondrites (Ruzicka et al. 2007; Kita et al. 2010). The trend for carbonaceous chondrites is generally attributed to an oxidizing agent comprising $^{16}$O-depleted ice exchanging with originally $^{16}$O-enriched silicates in the formation regions of type II chondrules (e.g. Tenner et al. 2013). For ordinary chondrites, the disappearance of the trend would mean, in this picture, that the "reduced endmember" had roughly the same isotopic composition as the oxidizing agent. This may be possibly due to the loss, relative to carbonaceous chondrites, of $^{16}$O-rich components such as refractory inclusions (see e.g. Zanda et al. 2006, Jacquet 2014b) and/or extensive processing and mixing of an originally carbonaceous chondrite-like "reduced endmember" with the oxidizing agent. In the latter scenario, the said oxidizing agent could have undergone Fe reduction and loss, hereby explaining the lower Fe/Mn ratio of type II chondrule olivine in ordinary chondrites compared to carbonaceous chondrites (Berlin et al. 2011); this difference could alternatively reflect the less moderately volatile element-depleted composition of ordinary chondrites compared to carbonaceous ones.

## *5. Conclusions*

We have performed LA-ICP-MS of silicate phases in type I and type II chondrules in LL3 chondrites Bishunpur and Semarkona.

The results are fairly similar to those reported previously for CR and CV chondrites, with mesostasis having REE concentrations around 10 x CI, olivine and low-Ca pyroxene having HREE concentrations at 0.1-1 X CI, and LREE concentrations at 0.01-0.1 x CI typically (with olivine poorer in REE than pyroxene). Very LREE-depleted olivine as encountered before in carbonaceous chondrites is however lacking $((Ce/Yb)_N > 0.02)$. On average, type II chondrule olivine has lower (by a factor of ~2) incompatible lithophile element concentrations than its type I counterpart. Na is present in both type I and type II chondrule olivine.

Based on these data and on the literature, we have inferred the following differences between the formation processes of type I and type II chondrules:

1° Higher oxygen fugacities and higher compound chondrule frequency for type II chondrules suggest that solid concentrations were higher in the formation regions of the latter than those of type I chondrules.

2° Incompatible element (e.g. REE) systematics suggest that type I and type II chondrule olivine record batch and fractional crystallization, respectively, for most minor and trace elements. Type II chondrules thus appear to have undergone faster cooling rates than type I chondrules (except possibly during pyroxene crystallization in the latter), with a rough boundary between the two types at ~10 K/h (assuming a single-stage cooling). The overabundance of very incompatible elements (LREE), though probably a kinetic effect, should not be used to calibrate cooling rates.

3° Type II chondrules lost less moderately volatile elements (including Fe), at the end of the process. They in particular seem to have been closed systems for Na which may have been inherited from an albitic precursor, as debris of metamorphosed chondrites, or mesostases from previous generations of chondrules. Lack of substantial Na loss indicates that the ambient gas was far from canonical (regardless of devolatilization from chondrules), pointing perhaps to planetary neighborhoods as chondrule-forming regions.

These differences are incompatible with common chondrule-forming regions for the two types and point to distinct environments. These differences need to be unified in any valid chondrule formation model. While many of them may be readily linked from a cosmochemical perspective, a positive correlation between solid concentration and cooling rate emerges as a key remaining constraint for such a theory.

| Chondrule/phase | Type I olivine | | | Type II olivine | | |
|---|---|---|---|---|---|---|
| | Average | 1st quartile | 3rd quartile | Average | 1st quartile | 3rd quartile |
| Li | 0.703 | 0.614 | 0.992 | 0.758 | 0.637 | 1.039 |
| Na | 42.236 | 20.500 | 65.222 | 49.278 | 44.280 | 150.055 |
| Al | 466 | 304 | 901 | 170 | 130 | 239 |
| Si | 195721 | 192712 | 198575 | 187334.1 | 182301 | 189655 |
| P | 27.0 | 16.5 | 48.4 | 61 | 24.0 | 130.0 |
| Ca | 1095 | 666 | 2001 | 546 | 494 | 727 |
| Sc | 5.80 | 3.97 | 9.39 | 2.83 | 2.41 | 3.55 |
| Ti | 143 | 102 | 250 | 23 | 19 | 32 |
| V | 86 | 98 | 116 | 55 | 40 | 69 |
| Cr | 1901 | 1544 | 3227 | 2645 | 2059 | 3017 |
| Mn | 807 | 491 | 1169 | 2476 | 2091 | 2621 |
| Co | 4.93 | 3.49 | 7.76 | 26.8 | 12.33 | 47.06 |
| Ni | 27.6 | 19.9 | 40.4 | 58.0 | 27.8 | 92.2 |
| Cu | 0.357 | 0.170 | 0.697 | 0.29 | 0.155 | 0.488 |
| Zn | 3.68 | 2.24 | 4.50 | 4.357 | 3.14 | 5.85 |
| Sr | 0.200 | 0.119 | 0.330 | 0.060 | 0.040 | 0.092 |
| Y | 0.110 | 0.073 | 0.217 | 0.035 | 0.029 | 0.045 |
| Zr | 0.097 | 0.054 | 0.156 | 0.040 | 0.028 | 0.066 |
| Nb | 0.013 | 0.006 | 0.028 | 0.011 | 0.010 | 0.014 |
| Ba | 0.062 | 0.014 | 0.142 | 0.048 | 0.023 | 0.107 |
| La | 0.004 | 0.002 | 0.011 | 0.002 | 0.001 | 0.003 |
| Ce | 0.014 | 0.006 | 0.033 | 0.003 | 0.002 | 0.005 |
| Pr | 0.001 | 0.001 | 0.003 | 0.001 | 0.000 | 0.001 |
| Nd | 0.012 | 0.006 | 0.021 | 0.004 | 0.004 | 0.009 |
| Sm | 0.008 | 0.006 | 0.010 | 0.004 | 0.002 | 0.008 |
| Eu | 0.002 | 0.001 | 0.003 | 0.001 | 0.001 | 0.002 |
| Gd | 0.010 | 0.008 | 0.016 | 0.005 | 0.003 | 0.010 |
| Dy | 0.013 | 0.009 | 0.030 | 0.005 | 0.005 | 0.008 |
| Er | 0.015 | 0.009 | 0.023 | 0.005 | 0.004 | 0.008 |
| Yb | 0.027 | 0.018 | 0.047 | 0.009 | 0.008 | 0.013 |
| Lu | 0.005 | 0.004 | 0.010 | 0.003 | 0.002 | 0.003 |
| Hf | 0.002 | 0.002 | 0.002 | 0.002 | 0.002 | 0.002 |
| Pb | 0.039 | 0.016 | 0.085 | 0.041 | 0.030 | 0.078 |
| Th | 0.002 | 0.003 | 0.008 | 0.000 | 0.000 | 0.000 |
| U | 0.006 | 0.006 | 0.006 | 0.001 | 0.001 | 0.002 |
| Fe | 15608 | 9117 | 28665 | 98871 | 82278 | 132585 |

**Table 1: Summary of olivine compositions.**
All concentrations are expressed in ppm (and are drawn from LA-ICP-MS analyses except for Fe and Si known from EMPA). Averaging is geometrical. Smk24 and Bi7 are excluded from type I and type II averages.

| Chondrule/phase | Type I enstatite | | | Type I augite | Type II pyroxene | | |
|---|---|---|---|---|---|---|---|
| | Average | 1st quartile | 3rd quartile | Average | Average | 1st quartile | 3rd quartile |
| Li | 1.053 | 0.619 | 1.124 | 0.706 | 2.244 | 1.207 | 3.005 |
| Na | 459 | 361 | 462 | 7786 | 648 | 343 | 994 |
| Al | 3616 | 3565 | 5413 | 38103 | 2216 | 1678 | 2613 |
| Si | 273685 | 270495 | 276210 | 276285 | 263858 | 257419 | 265155 |
| P | 93 | 26.680 | 311.026 | 39.407 | 44 | 26.745 | 46.170 |
| Ca | 3571 | 3122 | 6236 | 124187 | 3958 | 2282 | 6516 |
| Sc | 14 | 13.2 | 19.7 | 61.5 | 10 | 6.7 | 13.7 |
| Ti | 675 | 666 | 1129 | 5206 | 276 | 180 | 357 |
| V | 102 | 88 | 118 | 191 | 118 | 86 | 146 |
| Cr | 4296 | 3881 | 5190 | 13118 | 6771 | 5486 | 7881 |
| Mn | 1211 | 805 | 1895 | 3468 | 4299 | 3523 | 4793 |
| Co | 4.6 | 1.40 | 10.39 | 2.32 | 16.6 | 13.59 | 26.90 |
| Ni | 57.58 | 14.9 | 85.8 | 23.5 | 49.53 | 44.9 | 98.0 |
| Cu | 4.02 | 3.02 | 5.37 | 11.33 | 2.21 | 1.04 | 3.38 |
| Zn | 9.261 | 3.15 | 19.43 | 15.55 | 4.453 | 1.89 | 7.35 |
| Sr | 0.924 | 0.707 | 1.258 | 12.773 | 0.777 | 0.359 | 1.497 |
| Y | 0.222 | 0.145 | 0.523 | 16.337 | 0.216 | 0.130 | 0.316 |
| Zr | 0.441 | 0.313 | 0.674 | 22.100 | 0.272 | 0.090 | 0.429 |
| Nb | 0.052 | 0.035 | 0.094 | 0.946 | 0.042 | 0.025 | 0.050 |
| Ba | 0.389 | 0.233 | 0.969 | 2.788 | 0.373 | 0.145 | 0.608 |
| La | 0.019 | 0.017 | 0.049 | 0.918 | 0.008 | 0.002 | 0.020 |
| Ce | 0.037 | 0.022 | 0.097 | 3.078 | 0.030 | 0.010 | 0.058 |
| Pr | 0.005 | 0.003 | 0.017 | 0.563 | 0.005 | 0.002 | 0.009 |
| Nd | 0.053 | 0.031 | 0.102 | 3.696 | 0.030 | 0.013 | 0.049 |
| Sm | 0.011 | 0.006 | 0.024 | 1.793 | 0.009 | 0.007 | 0.018 |
| Eu | 0.006 | 0.005 | 0.009 | 0.112 | 0.005 | 0.003 | 0.007 |
| Gd | 0.015 | 0.012 | 0.036 | 2.597 | 0.016 | 0.006 | 0.028 |
| Dy | 0.032 | 0.026 | 0.108 | 3.950 | 0.025 | 0.011 | 0.045 |
| Er | 0.030 | 0.017 | 0.078 | 2.839 | 0.021 | 0.015 | 0.036 |
| Yb | 0.050 | 0.034 | 0.091 | 1.997 | 0.036 | 0.018 | 0.054 |
| Lu | 0.009 | 0.008 | 0.021 | 0.357 | 0.006 | 0.004 | 0.008 |
| Hf | 0.022 | 0.019 | 0.031 | 0.745 | 0.004 | 0.004 | 0.004 |
| Pb | 0.197 | 0.064 | 0.301 | 0.501 | 0.125 | 0.046 | 0.245 |
| Th | 0.004 | 0.004 | 0.004 | 0.113 | 0.001 | 0.001 | 0.002 |
| U | 0.003 | 0.002 | 0.003 | 0.032 | 0.001 | 0.001 | 0.002 |
| Fe | 11288 | 6457 | 13817 | 15214 | 101617 | 103905 | 114137 |

**Table 2: Summary of pyroxene compositions.**
All concentrations are expressed in ppm (and are drawn from LA-ICP-MS analyses except for Fe and Si known from EMPA). Averaging is geometrical.

| Chondrule/phase | Type I mesostasis | | | Type II mesostasis | | |
| --- | --- | --- | --- | --- | --- | --- |
| | Average | 1st quartile | 3rd quartile | Average | 1st quartile | 3rd quartile |
| Li | 0.263 | 0.115 | 0.350 | 1.088 | 0.261 | 2.150 |
| Na | 27813 | 14996 | 23626 | 49835 | 39978 | 58308 |
| Al | 103405 | 96378 | 114106 | 70070 | 59836 | 76730 |
| Si | 259994 | 238546 | 280464 | 290090 | 280008 | 316877 |
| P | 853.863 | 45.444 | 143.613 | 2418.775 | 258.824 | 2859.836 |
| Ca | 73009 | 61520 | 92471 | 39563 | 26063 | 48554 |
| Sc | 32.2 | 26.6 | 40.4 | 18.9 | 10.9 | 26.3 |
| Ti | 4214 | 3678 | 4773 | 3802 | 3115 | 3715 |
| V | 39.9 | 18.1 | 44.7 | 48.3 | 15.7 | 65.5 |
| Cr | 3881 | 3137 | 4814 | 2299 | 735 | 2718 |
| Mn | 2155 | 680 | 2886 | 2763 | 1679 | 3685 |
| Co | 10.4 | 2.7 | 14.2 | 25.8 | 8.2 | 41.5 |
| Ni | 431 | 39 | 188 | 187 | 68 | 148 |
| Cu | 23.4 | 3.4 | 24.5 | 8.5 | 4.0 | 11.0 |
| Zn | 29.0 | 11.5 | 28.4 | 14.8 | 6.9 | 18.1 |
| Sr | 77.9 | 66.4 | 98.0 | 33.2 | 20.3 | 39.5 |
| Y | 14.0 | 13.2 | 14.8 | 11.0 | 9.6 | 11.6 |
| Zr | 38.0 | 37.4 | 41.7 | 30.3 | 23.7 | 30.6 |
| Nb | 2.906 | 2.240 | 4.126 | 2.886 | 2.228 | 2.990 |
| Ba | 17.5 | 10.7 | 27.5 | 23.6 | 9.2 | 14.6 |
| La | 2.681 | 2.280 | 3.183 | 2.426 | 1.789 | 2.694 |
| Ce | 7.530 | 6.450 | 8.681 | 6.681 | 5.013 | 7.341 |
| Pr | 1.081 | 1.006 | 1.250 | 0.925 | 0.673 | 1.023 |
| Nd | 5.300 | 4.910 | 6.085 | 4.649 | 3.603 | 5.078 |
| Sm | 1.730 | 1.630 | 1.993 | 1.329 | 0.993 | 1.513 |
| Eu | 0.549 | 0.477 | 0.673 | 0.418 | 0.202 | 0.543 |
| Gd | 1.919 | 1.794 | 2.074 | 1.674 | 1.390 | 1.743 |
| Dy | 2.666 | 2.315 | 2.940 | 2.112 | 1.666 | 2.386 |
| Er | 1.661 | 1.373 | 1.868 | 1.302 | 1.047 | 1.470 |
| Yb | 1.751 | 1.638 | 1.869 | 1.340 | 1.223 | 1.412 |
| Lu | 0.254 | 0.217 | 0.282 | 0.209 | 0.183 | 0.227 |
| Hf | 1.107 | 0.810 | 1.272 | 0.840 | 0.612 | 0.886 |
| Pb | 0.889 | 0.366 | 0.787 | 0.789 | 0.704 | 0.902 |
| Th | 0.372 | 0.310 | 0.456 | 0.280 | 0.210 | 0.287 |
| U | 0.080 | 0.052 | 0.095 | 0.086 | 0.059 | 0.093 |
| Fe | 8120 | 1022 | 12775 | 40803 | 27004 | 54902 |

**Table 3: Summary of mesostasis compositions.**
All concentrations are expressed in ppm (and are drawn from LA-ICP-MS analyses except for Fe and Si known from EMPA).

*Acknowledgments:* We are grateful to the Programme National de Planétologie and the Canadian Institute for Theoretical Astrophysics for funding this research. We thank the Muséum National d'Histoire Naturelle for the lease of samples. Reviews by Guy Libourel and Alan Rubin, as well as the Associate Editor Alexander Krot, allowed to significantly improve the focus and the discussion of alternative thermal histories.